\documentclass[DIV12,11pt]{scrartcl}
\usepackage[utf8]{inputenc}
\usepackage[english]{babel}

\usepackage{amsthm,amssymb,amsmath}
\usepackage[noblocks]{authblk}

\usepackage{multirow}
\usepackage{multicol}
\usepackage{graphicx}
\usepackage{algorithmic}
\usepackage{comment}
\usepackage{caption}
\usepackage{color}
\usepackage{setspace}
\usepackage{array}
\newcolumntype{L}[1]{>{\raggedright\let\newline\\\arraybackslash\hspace{0pt}}m{#1}}
\newcolumntype{C}[1]{>{\centering\let\newline\\\arraybackslash\hspace{0pt}}m{#1}}
\newcolumntype{R}[1]{>{\raggedleft\let\newline\\\arraybackslash\hspace{0pt}}m{#1}}
\usepackage{tabulary}

\setkomafont{title}{\normalfont \LARGE \bfseries}
\setkomafont{section}{\normalfont \Large \bfseries \boldmath}
\setkomafont{subsection}{\normalfont \large \bfseries}
\setkomafont{subsubsection}{\normalfont \normalsize \bfseries}
\setkomafont{descriptionlabel}{\itshape}
\setkomafont{paragraph}{\normalfont \bfseries \boldmath}

\newtheorem{theorem}{Theorem}

\newtheorem{proposition}[theorem]{Proposition}

\usepackage{fancyhdr}
\fancypagestyle{myplain}
{
  \fancyhf{}

  \fancyfoot[C]{\thepage}
}
\title{A Branch--Price--and--Cut Algorithm for Optimal Decoding of LDPC Codes} % with High Error Correction Capability  in Communication Systems
\vspace{-3mm}
\author[1]{Banu Kabakulak\thanks{Corresponding author.  Tel.: +90 2123596771; fax: +90 2122651800. \\ E-mail addresses: banu.kabakulak@boun.edu.tr (B. Kabakulak), caner.taskin@boun.edu.tr (Z. C. Ta\c{s}k\i n), ali.pusane@boun.edu.tr (A. E. Pusane).}} 
\author[1]{Z. Caner Ta\c{s}k\i n}
\author[2]{Ali Emre Pusane}
\vspace{-3mm}
\affil[1]{Department of Industrial Engineering, Bo\u{g}azi\c{c}i University, \.{I}stanbul, Turkey}
\affil[2]{Department of Electrical and Electronics Engineering, Bo\u{g}azi\c{c}i University, \.{I}stanbul, Turkey}

\date{\vspace*{-2em}}

\begin{document}

\maketitle

\thispagestyle{empty}

\begin{abstract}

\vspace{-8mm}

Channel coding aims to minimize errors that occur during the transmission of digital information from one place to another. Low--density parity--check (LDPC) codes can detect and correct transmission errors if one encodes the original information by adding redundant bits. In practice, heuristic iterative decoding algorithms are used to decode the received vector. However, these algorithms may fail to decode if the received vector contains multiple errors. We consider decoding the received vector with minimum error as an integer programming problem and propose a branch--and--price method for its solution. We improve the performance of our method by introducing heuristic feasible solutions and adding valid cuts to the mathematical formulation. Computational results reveal that our branch--price--and--cut algorithm significantly improves solvability of the problem compared to a commercial solver in high channel error rates. Our proposed algorithm can find higher quality solutions than commonly used iterative decoding heuristics. %cannot find near--optimal feasible solutions or  branch--and--price
%\vspace{-3mm}

\textbf{Keywords:} Telecommunications, LDPC decoding, integer programming, branch--price--and--cut algorithm.

\end{abstract}

\section{Introduction and Literature Review} \label{Introduction and Literature Review}

Low--density parity--check (LDPC) codes, a category of linear block codes, were first investigated by Gallager \cite{G62} and rediscovered by MacKay \cite{M97, M99}. LDPC codes are now being used in hard disk drive read channels, Wireless (IEEE 802.11n/ IEEE 802.11ac, IEEE 802.16e WiMax), 10-GB, DVB-S2, and more recently in Flash SSD due to their high error detection and correction capabilities \cite{KOS14}. LDPC codes can be represented as sparse bipartite graphs known as Tanner graphs \cite{RL09}. Sparsity of Tanner graphs allows implementation of iterative message--passing decoding algorithms easily with low complexity and low decoding latency. Among message--passing algorithms, Gallager A and B algorithms and sum product (also known as belief propagation) algorithm are popular \cite{G62}, \ \cite{RU01} -- \cite{T81}. There are algorithms in literature that aim to reduce the complexity of the sum product (SP) algorithm \cite{FMI99} -- \cite{SLE14}.

Maximum likelihood (ML) decoding aims to decode a received vector by explicitly minimizing error probability. The ML decoding problem is NP--hard \cite{BMT78}. Hence, iterative message--passing algorithms are used in practice, although they are heuristic approaches. They can give close results to ML decoding on sparse Tanner graphs \cite{L05}. However, these algorithms do not guarantee optimality of the decoded vector, and they may fail to decode correctly when Tanner graph includes cycles.

ML decoding problem can be represented as an integer programming (IP) problem (given as EM formulation in Section \ref{MathematicalFormulations}). In 
\cite{FWK05}, linear relaxation of the IP problem is alternatively formulated (given as LPM formulation in Section \ref{MathematicalFormulations}). The authors utilize optimization techniques on a linear programming (LP) formulation and develop LP decoding algorithm for ML decoding problem. In \cite{VK07}, an iterative approach similar to SP is implemented for low complexity LP decoding and the technique is improved in \cite{B09}. The vertices of LP formulation are known as pseudocodewords. An efficent pseudocodeword search algorithm for LP decoding is introduced in \cite{CS08}. In \cite{TS08}, an LP decoding algorithm that adds necessary constraints as needed to the LP formulation is developed. The authors also include some valid inequalities introducing redundant check nodes. This LP decoder is further improved by separating pseudocodewords with new cuts in \cite{ZS12}. 

EM formulation is reformulated in \cite{YWF08} with fewer constraints. The authors solve the new formulation with a branch--and--bound algorithm. LPM formulation is addressed in \cite{TRH+10}, where the authors propose a separation algorithm to improve the error correction capability of LP decoder. Lagrangean relaxation techniques are applied to LPM formulation in \cite{ZS13, BLD+13}. %However, the proposed models do not allow decoding in an acceptable amount of time for codes with practical lengths. 

In this study, we consider LPM formulation and develop a branch--price--and--cut (BPC) algorithm for solving ML decoding problem for practical code lengths (approximately  $n = 4000$) efficiently. The rest of the paper is organized as follows: we define the problem formally in the next section. Section \ref{SolutionMethods} explains the proposed decoding techniques. In particular, we give the details of our branch--and--price (BP) algorithm in Section \ref{BranchAndPriceAlgorithm} and improvements to BP algorithm towards BPC in Section \ref{Improvements}. We present the corresponding computational results in Section \ref{ComputationalResults}. Some concluding remarks and comments on future work appear in Section \ref{Conclusions}.

%In this study, we consider LDPC--C codes and propose optimization based sliding window decoders that can give a near optimal decoded codeword for a received vector of practical length  (approximately  $n = 4000$)  in an acceptable amount of time. The mathematical formulation and proposed decoding algorithms are explained in Section \ref{SolutionMethods}. Our proposed decoders can be used in a real--time reliable communication system since they have low decoding latency. Besides, they are applicable in settings such as deep--space communication system due to their high error correction capability. 

  % We give the details of our  Branch--and--Price (BP) algorithm in Section \ref{sec:Branch_and_Price}. We consider to improve the performance of BP algorithm with some existing and developed feasible solution generation techniques in Section \ref{sec:FeasibleSolution}.

\section{Problem Definition}\label{ProblemDefinition}

\begin{figure}[h]
\begin{center}
\includegraphics[width=0.5\columnwidth]{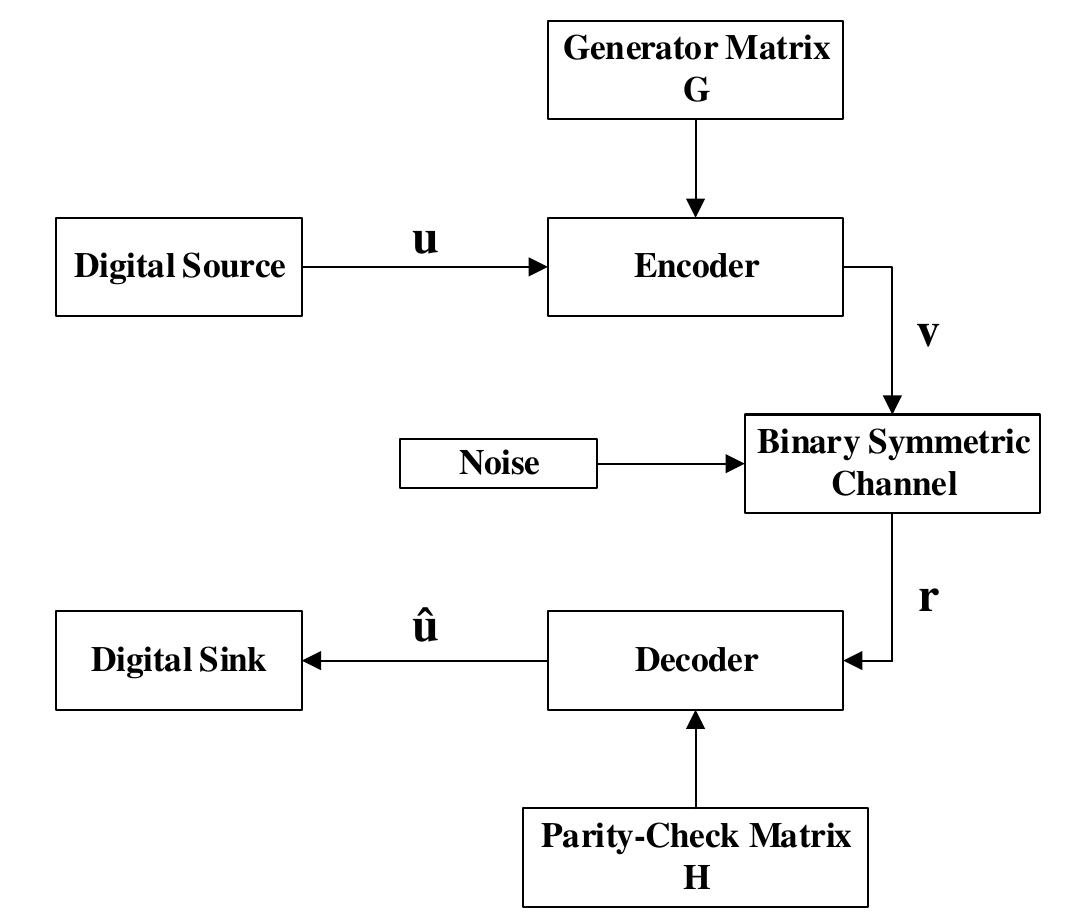}
\end{center}
\caption{Digital communication system diagram}
\label{fig:diagram}
\end{figure}

In a digital communication system, information is sent from a source to a sink over a noisy communication channel as shown in Figure \ref{fig:diagram}. We can represent original information as $k-$bits long sequence $\mathbf{u} = (u_{1}\ u_{2} \ ... \ u_{k})$  $(u_{i} \in \{0, 1\})$. Since there is noise in the communication channel, some bits of $\mathbf{u}$ can change. Information $\mathbf{u}$ is encoded with a $k \times n$ generator matrix $\mathbf{G}$ through the operation $\mathbf{v} = \mathbf{uG}$ (mod 2) to have more reliable communication. That is, $(n-k)$ redundant parity--check bits are added to $\mathbf{u}$ and $n$--bits long $(n \geq k)$ codeword $\mathbf{v} = (v_{1} \ v_{2} \ ... \ v_{n})$  $(v_{i} \in \{0, 1\})$ is obtained. 

The codeword $\mathbf{v}$ is transmitted over  a noisy communication channel to the receiver. In this paper we focus on Binary Symmetric Channel (BSC). In BSC, a bit $v_i$ is received correctly with probability $1-p$ or an error occurs with probability $p$  \cite{M03, HS14}.  The value of $v_i$ flips  in the case of an error, i.e., if $v_i = 0$, it becomes 1 and vice versa. Decoder tests the correctness of $n$--bits long received vector $\mathbf{r}$ with a $(n-k) \times n$  parity--check matrix $\mathbf{H}$. The received vector $\mathbf{r}$ is detected to be erroneous if $\mathbf{rH}^\textrm{T} \neq \mathbf{0}$ (mod 2). In such a case, the decoder runs decoding algorithms to fix the errors and estimate the original information as $\hat{\mathbf{u}}$ \cite{M05}.  % Buraya dergi ref'i konabilir belki...

One can obtain a generator matrix $\mathbf{G}$, which is not necessarily unique, from parity--check matrix $\mathbf{H}$ by carrying out binary arithmetic \cite{M03}. A vector $\mathbf{v}$  is a codeword if $\mathbf{v}\mathbf{H}^\textrm{T}=\mathbf{0}$ (mod 2). For any ($\mathbf{G, H}$) pair $\mathbf{GH}^\textrm{T} = \mathbf{0}$ (mod 2) holds, meaning that each row of $\mathbf{G}$ is a codeword. Moreover, the codewords are in the null space of $\mathbf{H}$ matrix and $\mathbf{G}$ is a basis for the null space. 

\begin{figure}[h]
\begin{center}
\begin{equation*}
\mathbf{H}=\begin{bmatrix}
\mathbf{I}_s^{1}  & \mathbf{I}_s^{2}  & \mathbf{I}_s^{3}  & \mathbf{I}_s^{4}   & \mathbf{I}_s^{5}  & \mathbf{I}_s^{6}  \\
 \mathbf{I}_s^{7}  & \mathbf{I}_s^{8}   & \mathbf{I}_s^{9} & \mathbf{I}_s ^{10} & \mathbf{I}_s^{11}  & \mathbf{I}_s^{12}   \\
 \mathbf{I}_s^{13}  & \mathbf{I}_s^{14}   & \mathbf{I}_s^{15}  & \mathbf{I}_s^{16}  & \mathbf{I}_s^{17}  & \mathbf{I}_s^{18}  \\
\end{bmatrix} \ \ \
\mathbf{H}=\begin{bmatrix}
1 \ 0 \ 0 \ 1 \ 0 \ 1 \ 1 \ 0 \ 0 \ 1 \ 1 \ 0 \\
0 \ 1 \ 1 \ 0 \ 1 \ 0 \ 0 \ 1 \ 1 \ 0 \ 0 \ 1 \\
0 \ 1 \ 1 \ 0 \ 0 \ 1 \ 1 \ 0 \ 0 \ 1 \ 0 \ 1 \\
1 \ 0 \ 0 \ 1 \ 1 \ 0 \ 0 \ 1 \ 1 \ 0 \ 1 \ 0 \\
0 \ 1 \ 0 \ 1 \ 1 \ 0 \ 0 \ 1 \ 1 \ 0 \ 0 \ 1 \\ 
1 \ 0 \ 1 \ 0 \ 0 \ 1 \ 1 \ 0 \ 0 \ 1 \ 1 \ 0 \\
\end{bmatrix}
\end{equation*}
(a) \hspace{50mm} (b) \vspace{-3mm}
\end{center}
\caption{(3, 6)--regular permutation matrix}
\label{fig:HMatrix}
\end{figure}
\vspace{-3mm}

$(J, K)-$regular LDPC codes are members of linear block codes that can be represented by a parity--check matrix $\mathbf{H}$ having $J-$many ones at each column and $K-$many ones at each row. One can generate a $(J, K)-$regular $\mathbf{H}$ matrix of dimension $(J s, K s)$ by randomly permuting the columns of an $s \times s$ identity matrix $ \mathbf{I}_s$. Regularity of the matrix is provided through augmenting identity matrices $K$ times at each row and $J$ times at each column. The generic structure of a $(3, 6)-$regular $\mathbf{H}$ matrix is given in Figure \ref{fig:HMatrix}a where $ \mathbf{I}_s^{i}$ represents the $i$th randomly permuted identity matrix. We give an example of a $(3, 6)-$regular $\mathbf{H}$ matrix with $s = 2$ in Figure \ref{fig:HMatrix}b.

\begin{figure}[h]
\begin{center}
\includegraphics[width=0.80\columnwidth]{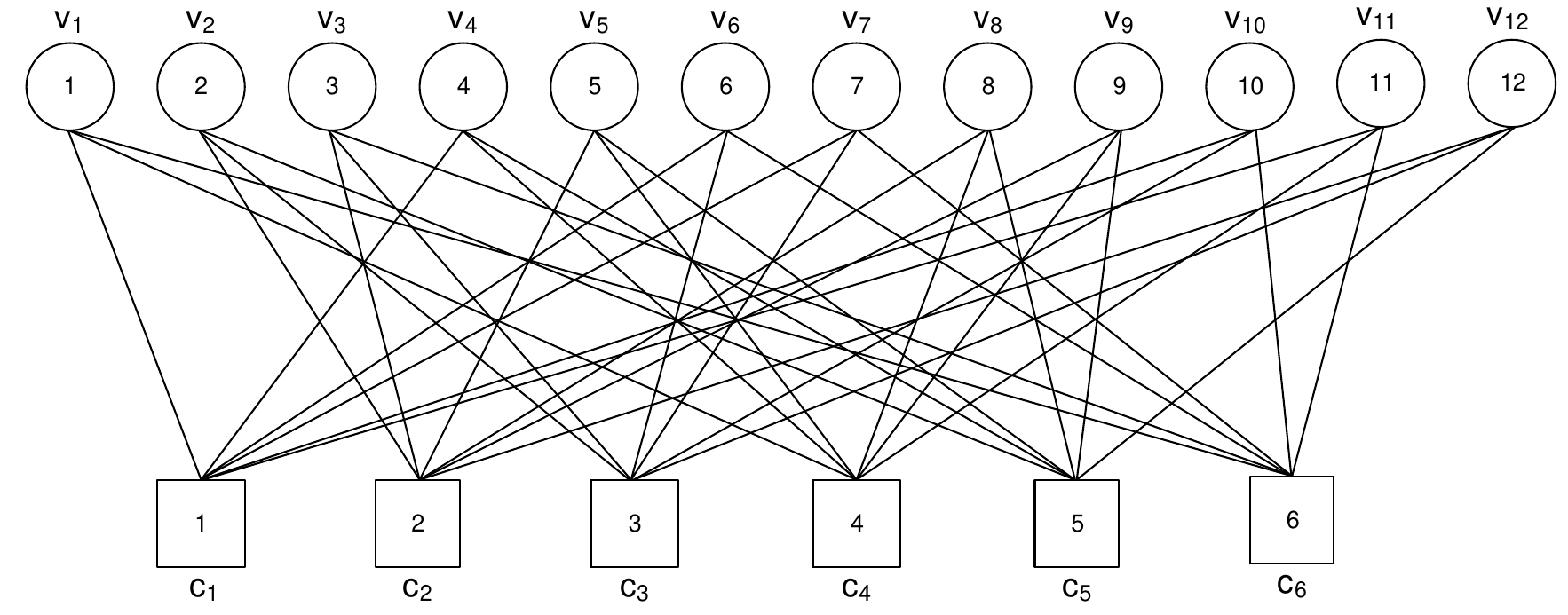}
\end{center}
\caption{Tanner graph representation of $\mathbf{H}$ matrix in Figure \ref{fig:HMatrix}b}
\label{fig:TannerGraph}
\end{figure}
\vspace{-3mm}

An LDPC code can alternatively be represented as a Tanner graph, which is a sparse bipartite graph, corresponding to $\mathbf{H}$ matrix  \cite{T81}. On one part of Tanner graph there is a variable node $i$ ($v_{i}$), $i \in \{1, ..., n\}$, for each bit of received vector. Each row of $\mathbf{H}$ matrix represents a parity--check equation and corresponds to a check node $j$ ($c_{j}), j \in \{1, ..., n-k\}$, in the other part of Tanner graph. A check node is said to be satisfied if its parity--check equation is equal to zero in (mod 2). The set of adjacent check (variable) nodes to a variable node $i$ (check node $j$) is represented by $N(v_i) (N(c_j))$. The \emph{degree} of $v_{i}$ ($c_{j}$) is the number of adjacent check nodes (variable nodes) on Tanner graph. That is, degree of $v_i$ is $d_i = |N(v_i)|$ and $c_j$ is $d_j = |N(c_j)|$.  Hence, $\mathbf{H}$ matrix is the bi--adjacency matrix of Tanner graph. Figure \ref{fig:TannerGraph} shows Tanner graph representation of the $\mathbf{H}$ matrix defined in Figure \ref{fig:HMatrix}b. %This representation of LDPC codes is practical due to the advantage of applying iterative decoding algorithms easily. 

In practical applications, iterative mesage--passing algorithms (such as Gallager A and SP) decode received vector $\mathbf{r}$ on Tanner graph efficiently due to its sparsity property \cite{L05}. However, these algorithms are heuristic approaches and they cannot guarantee that the solution is near optimal. As we show with computational experiments in Table \ref{tab:GallagerA} of Section \ref{ComputationalResults} for Gallager A, their error correction capability decreases significantly as the error probability increases.  Besides, they may fail to decode if the received vector includes multiple errors.

In this study, we focus on developing ML decoding algorithms using optimization techniques. In particular, we make use of a mathematical formulation by Feldman \emph{et al.} \cite{FWK05} for our BP algorithm. Then, we propose improvements to our BP algorithm to evolve our final BPC algorithm. %branch--and--price (BP) branch--price--and--cut (BPC)

\section{Solution Methods}\label{SolutionMethods}

We propose a BP algorithm (explained in Section \ref{BranchAndPriceAlgorithm}) for LDPC decoding problem. We improve the performance of BP method by providing feasible solutions via random sum heuristic (branch--and--price--random--sum (BPRS) method explained in Section \ref{RandomSum}) and tightening node relaxations with valid cuts (BPC method explained in Section \ref{ValidInequalities}). The terminology used in this paper is summarized in Table \ref{tab:lop}.

\begin{onehalfspace}
\footnotesize
\begin{center}
\captionof{table}{List of symbols}
    \label{tab:lop}
\begin{tabular}{l l l l} %p{6cm}}
\hline
\multicolumn{4}{c}{\textit{Parameters}} \\
\cline{1-4}
$C$ & set of check nodes & $k$  & length of the original information   \\  
$c_j$ & check node $j$ & $n$  & length of the encoded information, $|V|$ \\
$V$ & set of variable nodes & & number of columns in $\mathbf{H}$ \\
$v_i$ & variable node $i$ & $m$ & $n-k$, $|C|$, number of rows in $\mathbf{H}$ \\
$d_j (d_i)$ & degree of $c_j (v_i)$ in Tanner graph & $p$  & error probability in BSC   \\
$N(c_j) (N(v_i))$ & set of variable (check) nodes adjacent to $c_j (v_i)$ &  $\mathbf{u}$ & original information \\
$\mathbf{G}$ & generator matrix & $\mathbf{v}$ & encoded information\\
$\mathbf{H}$ & parity--check matrix & $\mathbf{r}$ & received vector \\
$\varepsilon_j$ & set of feasible local codewords for $c_j$ & $t_{max}$ & number of trials in RS heuristic\\
$\gamma_i$ & log--likelihood ratio for bit $i$ \\
\hline
\multicolumn{4}{c}{\textit{Decision Variables}} \\
\cline{1-4}
$f_i$ & $i$th bit of the decoded vector &  $\mu_j$  &  dual variable for constraints (\ref{cons3}) \\
 $w_{jS}$ & 1 if local codeword $S$ of $c_j$ is selected, & \multirow{2}{*}{$\tau_{ij}$} & \multirow{2}{*}{dual variable for constraints (\ref{cons4})} \\
& 0 otherwise \\
$l_j$ & an auxiliary integer variable &  $\zeta_j$ & optimum objective function value \\
$x_i $ & 1 if $i \in S$ of $c_j$, 0 otherwise & & of Subproblem$(j)$ \\
\hline
\end{tabular}
\end{center}
\end{onehalfspace}

\subsection{Mathematical Formulations} \label{MathematicalFormulations}

%In this section, we summarize our work on designing a decoding algorithm with high error correction capability for LDPC codes. 

The decoding problem can be represented with Exact Model (EM), which is  given in \cite{KD10}. Columns and rows of a $(n-k) \times n$ parity--check matrix $\mathbf{H}$ of a binary linear code can be represented with index sets $V = \{1, ..., n\}$ and $C = \{1, ..., n-k\}$, respectively.
In EM, $H_{ji}$ is the $(j, i)-$entry of parity--check matrix $\mathbf{H}$, $f_i$ is a binary variable denoting the value of the $i$th code bit and $l_j$ is an integer variable. Here, $\mathbf{r}$ represents the received vector. \\ 

\textbf{Exact Model (EM):}
\begin{align}
% \hspace{-3pt} \mbox{\textbf{EM:}} \hspace{11pt} & \min\hspace{5pt} \sum_{i: \hat{y}_i=1}(1- f_i) + \sum_{i: \hat{y}_i=0} f_i \label{Hamming}\\
& \hspace{-3pt}  \min\hspace{5pt} \sum_{i: r_i=1}(1- f_i) + \sum_{i: r_i=0} f_i  \label{Hamming}\\
&\hspace{25pt}\mbox{s.t.} \nonumber  \\  
&\hspace{25pt}\sum_{i \in V}H_{ji}f_i = 2l_j,  \hspace{9mm} \forall j \in C \label{sumtoeven}\\
&\hspace{25pt} f_i \in \{0, 1\},  \hspace{18mm} \forall i \in V, \label{f_vars}\\ 
&\hspace{25pt}  l_j \geq 0, \ l_j \in  \mathbb{Z},  \hspace{13mm} \forall j \in C. \label{k_vars}
\end{align}

Constraints (\ref{sumtoeven}) guarantee that the decoded vector $\mathbf{f}$ satisfies the equality $\mathbf{f} \mathbf{H}^\textrm{T}=\mathbf{0} \ \text{(mod 2)}$. The objective (\ref{Hamming}) minimizes the Hamming distance between the decoded vector $\mathbf{f}$ and the received vector $\mathbf{r}$. Hamming distance counts the number of different entries among two vectors. That is, the aim is to find the nearest codeword to the received vector. Constraints (\ref{f_vars}) and (\ref{k_vars}) set the binary and integrality restrictions on decision variables $\mathbf{f}$ and $\mathbf{k}$, respectively.

An alternative objective function is log--likelihood objective, which can be given as

\vspace{-5mm}

\begin{equation}
\min\hspace{5pt} \sum_{i \in V}  \gamma_{i}f_i  \label{log_like_obj}.
\end{equation}

Here, $\gamma_{i}$, as given in equation (\ref{log_like_value}), is a term that represents the log--likelihood ratio for received bit $i$. In this equation, $r_{i}$ represents the received value of bit $i$ and $f_{i}$ is the decoded value of the bit $i$. %error probability

\vspace{-10mm}

\begin{equation}
\gamma_{i}=log ( \tfrac{Pr(r_{i}\mid f_{i}=0)}{Pr(r_{i}\mid f_{i}=1)})  \label{log_like_value}
\end{equation}

As given in \cite{FWK05}, $\gamma_i = \log[p/(1-p)]$ if received bit $r_i = 1$ and $\gamma_i = \log[(1-p)/p]$ if $r_i = 0$ where $p$ is the error probability for BSC.

% HAMMING OBJECTIVE AND LOG__LIKELIHOOD OBJECTIVE ARE EQUIVALENT burada soylenebilir ins. Su anda Prunning section'ınında....

%As we explained in \textbf{Section \ref{sec:Introduction1}}, there are two alternative objective function definitions in literature for decoding problem. First is Hamming distance (given as equation (\ref{Hamming})) and second is log--likelihood  (given as equation (\ref{log_like_obj})) objectives. 

\begin{proposition}\label{prop5}
Hamming distance (equation (\ref{Hamming})) and Log--likelihood (equation (\ref{log_like_obj})) objectives are equivalent when $p < 0.5$. That is, both objectives give the same optimum solution set for the decoded codeword $\mathbf{f}$.
\end{proposition}

\emph{Proof.} First consider the log--likelihood objective.  The objective can be written as 

\vspace{-10mm}

\begin{align}
&\hspace{25pt} \min -\sum_{i: r_i=1} a f_i+ \sum_{i: r_i=0} a f_i \label{log_like}
\end{align}

where $a =  \log[(1-p)/p]$. Note that $a \geq 0$ for  $0 < p < 0.5$.  %We can assume that $p < 0.5$ without loss of generality, since It should be noted that without loss of generality it can be assumed that p<0.5, because if p>0.5 then all bits in y could simply be flipped first thereby replacing p with 1-p (and p=0.5 results in random data coming out of the channel...). Only after clearing up that p can be assumed to be smaller than 0.5 can be said that the two objectives are equivalent.

On the other hand, Hamming distance objective can be written as

\vspace{-8mm}

 \begin{align}
&\hspace{25pt} \min -\sum_{i: r_i=1} f_i + \sum_{i: r_i=0} f_i +  c_1 
\end{align}

where $c_1 = \sum_{i: r_i=1} 1$.

One can observe that Hamming distance objective is a scaled version of log--likelihood objective by choosing $a = 1$ and adding a constant term $c_1$.
Hence, both objectives have the same optimum solution set. $\square$\\

The linear relaxation of EM (LEM) can be obtained by replacing the constraints (\ref{f_vars}) and (\ref{k_vars}) with the following:

\vspace{-8mm}

 \begin{align}
&\hspace{25pt}  0 \leq f_i \leq 1, \  l_j \geq 0, \hspace{10mm} \forall i \in V, \ j \in C.
\end{align}

The same decoding problem is formulated as Integer Programming Master (IPM) in \cite{FWK05}, which  is a maximum likelihood decoder utilizing Tanner graph representation of $\mathbf{H}$ matrix.
%An alternative formulation for decoding problem is given in \cite{FMS+07} in order to find the nearest codeword to the received vector $\mathbf{\hat{y}}$. Integer Programming Master (IPM) formulation given in \cite{FMS+07} is a maximum likelihood decoder utilizing Tanner graph representation of $\mathbf{H}$ matrix. 
A \emph{local codeword} can be formed by assigning a value in \{0, 1\} to each variable node $i \in N(c_j)$ that is adjacent to $c_j$. A local codeword is \emph{feasible} if sum of the values of  variable nodes $i \in N(c_j)$ is zero in (mod 2).  For a check node $c_j$, the set of feasible local codewords can be given as $\varepsilon_{j} := \{S \subseteq N(c_j) : |S| \ even\}$. We can satisfy $c_j$ if we set each bit in $S$ to 1, and all other bits in $N(c_j)$ to 0. One can observe that $S = \emptyset$ trivially satisfies a check node, so  $\emptyset \in \varepsilon_{j}$ for all $c_j$. 

\begin{figure}[h]
\begin{center}
\includegraphics[width=0.50\columnwidth]{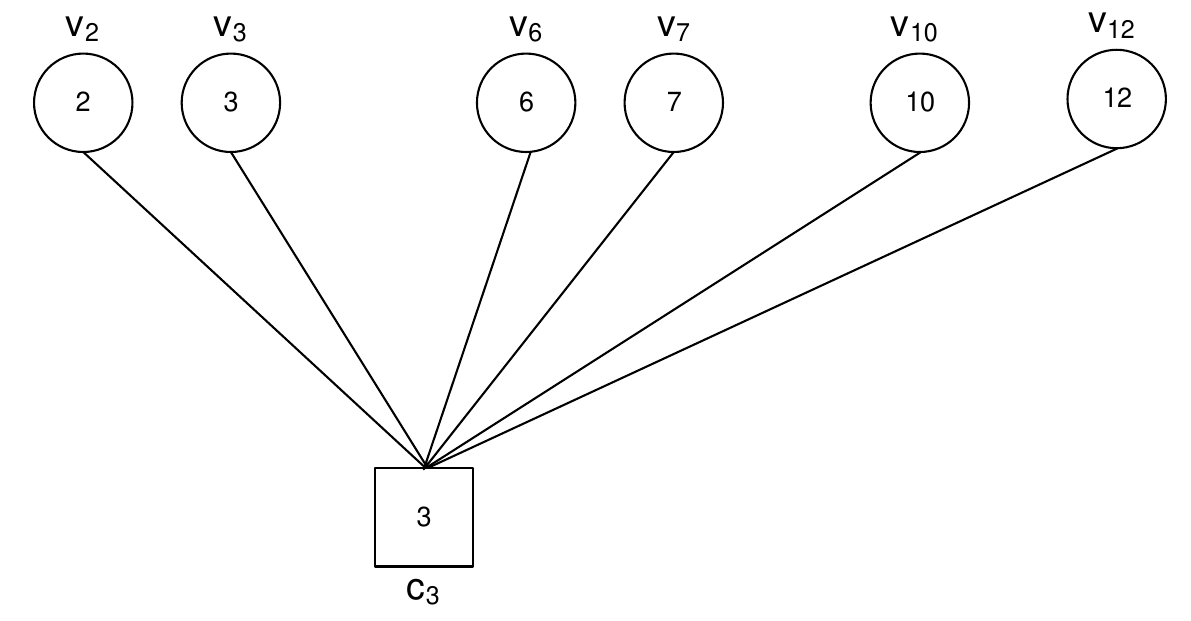}
\end{center}
\vspace{-3mm}
\caption{Neighbors of check node $c_3$ in Figure \ref{fig:TannerGraph}, $N(c_3)$}
\label{fig:LocalCodeword}

\end{figure}

In Figure \ref{fig:LocalCodeword}, we give the neighbors of check node  $c_3$ in Figure \ref{fig:TannerGraph} as an example. Parity--check equation for $c_3$ can be given as $c_3  = v_2 + v_3 + v_6 + v_7 + v_{10} + v_{12}$ (mod 2). We can see $c_3 = 0$ if we pick even number of neighboring variable nodes with value 1 and the remaining is 0. That is, $S = \{ 2, 6 ,7, 12\}$ is a feasible local codeword since  $c_3  = 1 + 0 + 1 + 1 + 0 + 1 = 0$ (mod 2). A \emph{codeword} is \{0, 1\} assignment of $v_i$ values for $i \in V$ that gives $c_j = 0$ for all $j \in C$. One can obtain a codeword by choosing a feasible local codeword for each $c_j$ that conforms with the feasible local codewords of other check nodes. For example, $(0 \ 1 \ 0 \ 0 \ 0 \ 1 \ 1 \ 1 \ 1 \ 0 \ 0 \ 1)$ is a codeword for Tanner graph in Figure 3.% observe that a codeword combines feasible local codewords of $c_j$s that are conforming with each other. 

\textbf{Integer Programming Master (IPM):}
%\vspace{-15pt}
\begin{align}
&\hspace{-3pt}\min\hspace{5pt} \sum_{i \in V}  \gamma_{i}f_i  \\ %\label{log_like_obj}\\
&\hspace{25pt}\mbox{s.t.} \nonumber  \\  
&\hspace{25pt}\sum_{S \in \varepsilon_{j}}w_{jS} = 1,  \hspace{35mm} \forall j \in C \label{cons3}\\
&\hspace{25pt}f_i - \sum_{S \in \varepsilon_{j} : i \in S}w_{jS} = 0, \hspace{23mm} \forall \  \text{edge} \ (i, j) \label{cons4}\\
&\hspace{25pt} f_i \geq 0, \ \forall i \in V, \ \  w_{jS} \in \{0, 1\}, \hspace{10mm} \forall j \in C, \forall S \in \varepsilon_{j}  \label{integer}.
\end{align}

In IPM model, binary decision variable $w_{jS}$ takes value 1 if feasible local codeword  $S \in \varepsilon_{j}$ of check node $c_j$ is selected and zero otherwise. Hence, decision variables $\mathbf{w}$ represent a feasible solution of parity--check equations and $f_{i}$ variable represents the decoded value of bit $i$. We can obtain a trivial solution (an upper bound) of IPM with $w_{j\emptyset} = 1$ for all $j \in C$ and $f_i = 0$ for all $i \in V$. We obtain Linear Programming Master (LPM) model by relaxing the constraints (\ref{integer}) as 

\vspace{-15mm}

\begin{align}
&\hspace{25pt}  f_i \geq 0, \ \forall i \in V, \ \  w_{jS} \geq 0, \hspace{10mm} \forall j \in C, \forall S \in \varepsilon_{j}  \label{linear}.
\end{align}

\subsubsection{On the Strength of LP Relaxations}

Let $z_{EM}(\mathbf{f})$ be the objective function value of EM for vector $\mathbf{f}$, and $z_{EM} = \min_{\mathbf{f}}z_{EM}(\mathbf{f})$ is the optimum objective function value. We observe that since EM objective is Hamming distance, $z_{EM} \geq 0$ for all $\mathbf{H}$ instances.
%We first observe that since EM objective is Hamming distance, it is nonnegative for all $\mathbf{H}$ instances, i.e., $z_{EM} \geq 0$ where $z_{EM} = \min_{\mathbf{f}}z_{EM}(\mathbf{f})$.
%We first observe that the objective function of EM minimizes the Hamming distance to the received codeword $\mathbf{\hat{y}}$. Then, the optimum function value of EM is nonnegative for all instances, i.e. $z_{EM} \geq 0$. 
 We have $z_{EM}(\mathbf{f}) = 0$ if $\mathbf{f = r}$, and for any feasible solution $\mathbf{f \neq r}$ the objective function value $z_{EM}(\mathbf{f}) > 0$. 

\begin{proposition}\label{prop8}
The optimum objective function value of LEM is 0 for all $\mathbf{H}$ instances, i.e., $z_{LEM} = 0$.
\end{proposition}

\emph{Proof.} %Let $\mathbf{f}$ be a fractional solution of LEM. Then, there is $i \in V$ such that $0 < f_i < 1$. If $r_i = 1$, then we will have cost $(1 - f_i) > 0$ and if $r_i = 0$, then we will have cost $f_i> 0$ to be added to the objective function. Then, for any fractional solution we have $z_{LEM}(\mathbf{f}) > 0$. In general if $\mathbf{f \neq r}$, then $z_{LEM}(\mathbf{f}) > 0$. 
%Let $\mathbf{f = r}$, then $\mathbf{f}$ is feasible for LEM since $0 \leq f_i \leq 1 \ \forall i$ and $k_j = \frac{\sum_{i \in V}H_{ij}f_i}{2} \geq 0$ since $H_{ji}$ is a matrix of 0s and 1s. 
Let $\mathbf{f = r}$ and $l_j = \frac{\sum_{i \in V}H_{ij}f_i}{2}$ for all $j$. Then $(\mathbf{f}, \mathbf{k})$ is feasible for LEM since $0 \leq f_i \leq 1 \ \forall i$ and $l_j  \geq 0$  for all $j$. % since $H_{ji}$ is a matrix of 0s and 1s. 
Then, the optimum objective function value $z_{LEM} = 0$ for all $\mathbf{H}$ instances. $\square$

\begin{proposition}\label{prop9}
LPM problem with Hamming distance objective (\ref{Hamming}) has strictly positive optimum objective value, i.e., $z_{LPM} > 0$, unless received codeword $\mathbf{r}$ is a feasible codeword.
\end{proposition}

\emph{Proof.} If received vector $\mathbf{r}$ is a feasible codeword, then  $\mathbf{f = r}$ be a feasible solution for LPM and it will be optimal. 
Assume that $\mathbf{r}$ is not a feasible codeword. Then, LPM problem has a fractional or integral feasible solution $\mathbf{f \neq r}$. This means that the Hamming distance objective is strictly positive for this optimal solution.
Hence, $z_{LPM} > 0$, if received codeword $\mathbf{r}$ is not a feasible codeword.   $\square$

To summarize, linear relaxation of EM formulation gives $z_{LEM} = 0$ for all $\mathbf{H}$ instances. The linear relaxation of IPM problem gives $z_{LPM} = 0$ if the received codeword $\mathbf{r}$ is a feasible codeword, otherwise $z_{LPM} > 0$. This means that LPM gives a better lower bound for IPM objective than LEM. 

Note that EM and IPM are integer programming formulations, and it is not practical to obtain an optimal decoding using a commercial solver for real--sized LDPC codes. Hence, we develop a branch--price--and--cut algorithm for IPM as explained in the following sections. 

\subsection{Branch--and--Price Algorithm} \label{BranchAndPriceAlgorithm}

In this section, we introduce a branch--and--price algorithm for IPM formulation given in \cite{FWK05} in order to find the nearest codeword to the received vector $\mathbf{r}$. We first define dual variables $\mu_j$ for constraints (\ref{cons3}) and $\tau_{ij}$ for constraints (\ref{cons4}) in LPM, and obtain Dual LPM (DLPM) model. 

\textbf{Dual  LPM (DLPM):}
%\vspace{-15pt}
\begin{align}
&\hspace{-3pt}\max\hspace{5pt} \sum_{j \in C}  \mu_j\\
&\hspace{25pt}\mbox{s.t.} \nonumber  \\  
&\hspace{25pt}\sum_{i \in S}\tau_{ij} \geq \mu_j,  \hspace{29mm} \forall j \in C, \  S \in \varepsilon_{j} \label{cons1}\\
&\hspace{25pt}\sum_{j \in N(v_i)}\tau_{ij} \leq \gamma_i, \hspace{25mm} \forall i \in V \label{cons1.2}\\ 
&\hspace{25pt} \mu_j  \ \text{free}, \ \forall j \in C,  \ \tau_{ij}\ \text{free}, \hspace{10mm} \forall \ \text{edges} \ (i,j).
\end{align}

We consider a Restricted LPM (RLPM) that has a limited number of columns corresponding to $w_{jS}$ variables. At each iteration of our column generation algorithm, we search for columns corresponding to variables $w_{jS}$ having positive reduced cost, i.e., $\mu_j - \sum_{i \in S}\tau_{ij} > 0$, and add them to RLPM. Such $w_{jS}$ columns are equivalent to the violated constraints from constraints (\ref{cons1}) in DLPM. 
%That is we look for $j$ and $S$ that have $\mu_j - \sum_{i \in S}\tau_{i,j} > 0$. 
If $\zeta_j = \max\{\mu_j - \sum_{i \in S}\tau_{ij} : S \in \varepsilon_j \} > 0$ for some $j$, then we add the column $\biggl[\begin{smallmatrix}
0\\ e_j \\ A_l
 \end{smallmatrix} \biggr]$ for variable $w_{jS}$. Here, ${e_j}$ is a $m-$column vector, that has a 1 at $j$th row and 0 otherwise, and $A_l$ is a $(\sum_{i=1}^n d_i)-$column vector which has $-1$ at $l$th row if $l$th edge is the edge $(i,j)$ with $i  \in S$. %If $\zeta_j = 0 \ \ \forall j$, then we are at optimum solution of LPM. 

Thus, at each iteration of column generation algorithm, we seek a local codeword $S$ for check node $c_j$ by solving the following subproblem for each $j$: 
 
$\mathbf{Subproblem(j):}$
\vspace{-10pt}
\begin{align}
&\hspace{-3pt}\min\hspace{5pt}\sum_{i \in N(c_j)}\tau_{ij}x_i - \mu_j\\
&\hspace{25pt}\mbox{s.t.} \nonumber  \\  
&\hspace{25pt} \sum_{i \in N(c_j)}x_i = 2l,  \label{cons6}\\
&\hspace{25pt} x_i \in \{0,1\}, l \in \mathbb{Z}^+.
\end{align}

Decision variable $x_i = 1$ if $i \in S$ and 0 otherwise. We can solve the $j$th subproblem to optimality with Algorithm 1, which runs in $\mathcal{O}(n \log n)$ time due to sorting step. % where $n$ is the number of variable nodes.  

\begin{center}
\singlespace
$
\begin{tabular}{ll}
\textbf{Algorithm 1:} (Solve Subproblem($j$)) \\
\hline
 \textbf{Input}  $\tau_{ij}$ values \\
\hline
%\vspace{-8mm}\\
1. Sort the $\tau_{ij}$ values  in nondecreasing order.\\  
\hspace{10pt} Let $\tau_{ij}^t$ be the $t$th smallest $\tau_{ij}$ value. \\
2. Set $x_i = 0 \ \ \forall i \in N(c_j)$, set $t = 1$. \\
3. \textbf{If} $\tau_{i_1,j}^t + \tau_{i_2,j}^{t+1} < 0$, \textbf{Then} set $x_{i_1}=x_{i_2}=1$, \textbf{Else} STOP. \\
4. $t \gets t+2$, go to Step 3. \\
\hline
\textbf{Output} Subproblem($j$) is solved.\\
\hline
\end{tabular}
$
\end{center}

As mentioned before, $w_{j\emptyset} = 1$ for all $j \in C$ is a feasible solution for LPM. 
%We observe that, empty set is a local codeword for each check node and if we set $w_{j, \emptyset} = 1$ for all $j \in C$, we have a feasible solution of LPM. 
Hence, for all $j \in C$ we can take $(j, \emptyset)$ columns for the starting RLPM problem. %The solution of this RLPM is $f^*_i = 0$ for all $i \in V$ and $ w^*_{j, \emptyset} = 1$ for all $j \in C$ which is integral but not necessarily optimal. 
We can solve LPM to optimality by introducing columns to RLPM until we have $\zeta_j = 0$ for all $j$. Since our ultimate goal is to solve IPM, we need to branch on decision variables if optimum solution of LPM is fractional. In the next section we discuss some alternative branching strategies. 

\subsubsection{Branching in BP Algorithm}

If we have a fractional optimal solution of LPM, then we have either $w_{jS}$ or $f_i$ variables fractional. Before determining a branching strategy, we will first prove the following proposition.

\begin{proposition}\label{prop1}
In LPM problem, $f_i$ values are integral for all $i$ if and only if $w_{jS}$ values are integral for all $(j,S)$.
\end{proposition}

\emph{Proof.} ($\Leftarrow$) Assume that $w_{jS}$ values are integral $\forall  (j,S)$. Constraints (\ref{cons4}) imply that $f_i$ values are integral $\forall  i$, since each $f_i$ is the sum of integer numbers. Besides, we observe that $w_{jS}$ values can be either 0 or 1, so do the $f_i$ values.

 ($\Rightarrow$) Assume for contradiction $f_i$ integral but $\exists (j, S)$ such that $w_{jS}$ is not integral.  By constraints (\ref{cons3}), we know $\sum_{S \in \varepsilon_{j}}w_{jS} = 1$. Hence, for at least two  $w_{jS}$ variables, say  $w_{j,S_1} = \alpha$ and  $w_{j,S_2} = \beta$ with $\alpha, \beta > 0$ and $\alpha + \beta \leq 1$, we have fractional values. Since $S_1 \neq S_2$, there exists $l \in S_2 \setminus S_1$ without loss of generality.  %$\forall  S$.

For variable node $l$ and check node $j$, we have the constraint $f_l =  \sum_{S \in \varepsilon_{j} : l \in S}w_{jS}$ for edge $(l, j)$. Edge $(l, j)$ exists, since $l \in S_2 \in \varepsilon_{j}$ which implies that $l \in N(c_j)$. We know that $l \not \in S_1$, meaning that $w_{j,S_1} = \alpha$ will not be in the sum. This means $f_l =  \sum_{S \in \varepsilon_{j} : l \in S}w_{j,S} \leq 1- w_{j,S_1} = 1-\alpha < 1$. Moreover, $w_{j,S_2}$ will be in the sum, since $l \in S_2$. This gives $f_l \geq w_{j,S_2} = \beta > 0$. As a result, $0 < f_l < 1$, i.e., $f_l$  is a fractional value. This contradicts with our assumption that $f_i$ values are all integral. Hence, we conclude that if  $f_i$ integral $\forall i$, then $w_{jS}$ values are also integral $\forall  (j,S)$.
Combining two results, we see that $f_i$ values are integral $\forall  i$ if and only if $w_{jS}$ values are integral $\forall  (j,S)$. $\square$

%\vspace{3mm}

As a result of this proposition, in order to have an integral solution to the LPM problem, it is sufficient to either branch on $w_{j,S}$ variables to have integral $w_{jS}$ values or branch on $f_i$ variables to have integral $f_i$ values. Having integral $w_{jS}$ values (or integral $f_i$ values) will guarantee that all decision variables are integral. 

%_______________________________-

%{\bf Branching on $f_i$ variables:}
\paragraph{Branching on $f_i$ variables:} %subsub

Assume that we solve the RLPM and find that for some $v_i$,  $f_i$ is fractional. Then, we consider to branch the problem by assigning $f_i = 0$ in one branch and $f_i = 1$ in the other branch. We continue to branch on the $\textbf{f}$ variables until we have an integral solution for LPM problem, which is a feasible solution of IPM.  %We continue to branch on the $\textbf{f}$ variables until we have an integral solution in RLPM. In that case, we have an integer feasible solution for LPM problem, which is a feasible solution of IPM. 

\begin{comment}
\begin{figure}[htbp]
\begin{center}
\includegraphics[width=0.6\columnwidth]{branch.png}
\end{center}
\caption{Branching strategy.}
\vskip\baselineskip 
\label{fig:branching}
\end{figure}
\end{comment}

In a branch, we can say that $f_{i}= 0$  for $i \in N_0$ and $f_{i} = 1$  for $i \in N_1$,  where $N_0 \cup N_1  \subseteq V$ and $N_0 \cap N_1 = \emptyset$.  In this branch, we have the following subproblem $j$: % is that we have some $f_i$ variables are set to 0 and some of them are set to 1. When we are at the $r$th level, we can say that $f_{i}= 0$  for $i \in N_0$ and $f_{i} = 1$  for $i \in N_1$,  where $N_0 \cup N_1 = \bar{V} \subseteq V$, $|\bar{V}| = r$ and $N_0 \cap N_1 = \emptyset$. In this branch, we have added the following constraints to the subproblem $j$:
\begin{comment}
\vspace{-8mm}
\begin{align}
&\hspace{25pt} x_{i} = 0, \ \text{if} \ i \in N(c_j) \cap N_0, \ \text{and} \ x_{i} = 1, \  \text{if} \ i \in N(c_j) \cap N_1. \label{cons11}
\end{align}
\end{comment}

$\mathbf{Subproblem(j) \ on \ a \ branch:}$
\begin{align}
&\hspace{-3pt}\min\hspace{5pt}\sum_{i \in N(c_j)}\tau_{ij}x_i - \mu_j \nonumber\\
&\hspace{25pt}\mbox{s.t.} \nonumber  \\  
&\hspace{25pt} \sum_{i \in N(c_j)}x_i = 2l, \nonumber\\
&\hspace{25pt} x_{i} = 0, \hspace{10mm} \text{if} \ i \in N(c_j) \cap N_0,  \nonumber\\
&\hspace{25pt} x_{i} = 1, \hspace{10mm}  \text{if} \ i \in N(c_j) \cap N_1, \nonumber\\
&\hspace{25pt} x_i \in \{0,1\}, l \in \mathbb{Z}^+.\nonumber
\end{align}

In order to solve Subproblem$(j)$, we eliminate the $x_{i}$ variables for $ i \in N(c_j) \cap N_0$ and we plug in $x_{i} = 1$ values for $ i \in N(c_j) \cap N_1$ to obtain an additional constant term from the corresponding $\tau_{ij}$ values. We can solve the remaining problem by applying Algorithm 2, modified Algorithm 1, given below. Algorithm 2 also runs in $\mathcal{O}(n \log n)$ time. % due to sorting step where $n$ is the number of variable nodes. 

\begin{center}
\singlespace
$
\begin{tabular}{ll}
\textbf{Algorithm 2:} (Solve Subproblem($j$) on a branch) \\
\hline
\vspace{0.5mm}
\textbf{Input:} Sets $N_0$ and $N_1$, where  $f_{i}= 0$  for $i \in N_0$ and $f_{i} = 1$  for $i \in N_1$.\\
\hline
%\vspace{-8mm}\\
0. Set  $x_{i} = 0$,  if $i \in N(c_j) \cap N_0$, and $x_{i} = 1$,  if $i \in N(c_j) \cap N_1$.\\
\hspace{10pt} Let $\mathcal{I}_j = N(c_j) \setminus (N_0 \cup N_1)$.\\
1. Sort the $\tau_{ij}$ values in nondecreasing order for $i \in \mathcal{I}_j $.\\ 
\hspace{10pt} Let $\tau_{ij}^t$ be the $t$th smallest $\tau_{ij}$ value. \\
2. Set $x_i = 0 \ \ \forall i \in \mathcal{I}_j$, set $t = 1$. \\
3. \textbf{If} $| N(c_j) \cap N_1|$ is even \\
4. \hspace{6pt} \textbf{Then} set $x_{i_1} = x_{i_2} = 1$ if $\tau_{i_1,j}^t + \tau_{i_2,j}^{t+1} < 0$, otherwise STOP.\\
5. \hspace{6pt} $t \gets t+2$, go to Step 4. \\
6. \textbf{Else}  set $x_{i} = 1$ for $\tau_{ij}^t$ \\
7. \hspace{6pt} \textbf{If} $\tau_{ij}^t < 0$, \textbf{Then} $t \gets t+1$ and go to Step 4, \textbf{Else} STOP.\\
8. \textbf{End If} \\
\hline
%\vspace{-8mm}\\
\textbf{Output:} A local codeword $S$ with objective value \\
\hspace{40pt} $\zeta_j =  \sum_{i \in \mathcal{I}_j}\tau_{ij}x_i + \sum_{i \in N(c_j) \cap N_1}\tau_{ij} - \mu_j$. \\
\hline
\end{tabular}
$
\end{center}

From the above analysis, we observe that branching on $f_i$ variables does not change the structure of the subproblems. On the other hand, branching on $w_{jS}$ variables affects the subproblem structure. As a result, we branch on $f_i$ variables since we can find an optimal solution of the subproblems in polynomial time.

\subsubsection{Repairing Infeasibility in Node Relaxations}

\begin{figure}[htbp]
\begin{center}
\includegraphics[width=0.35\columnwidth]{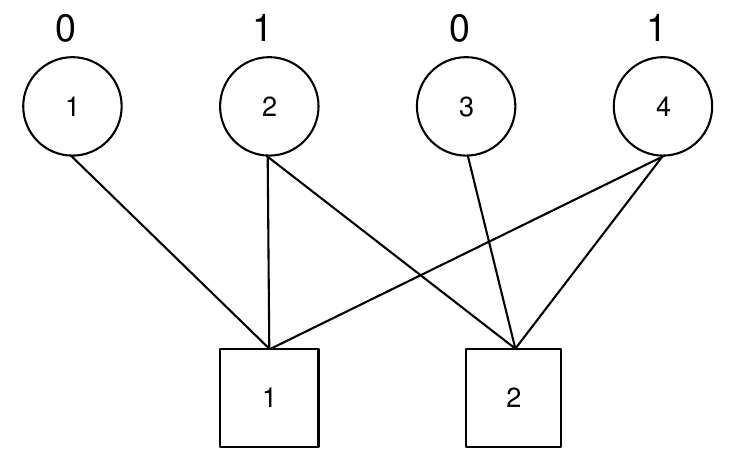}
\end{center}
\caption{An example Tanner graph}
\label{fig:ExampleTanner}
\end{figure}

In the application of BP algorithm, we observe that a branch can be pruned although there exists a feasible solution on that branch. This may happen if the currently generated columns are not sufficient to construct a feasible solution on the branch. 
As an example, consider we are at the $f_2 = 1$ and $f_4 = 1$ branch of Tanner graph in Figure \ref{fig:ExampleTanner}.

%\vspace{3mm}

The set of all feasible local codewords for check node $c_1$ is $\varepsilon_1 = \{\emptyset, \{1, 2\},$  $\{1, 4\}, \{2, 4\}\}$ and for check node $c_2$ is $\varepsilon_2 = \{\emptyset, \{2, 3\}, \{2, 4\}, \{3, 4\}\}$. On the $f_2 = 1$ and $f_4 = 1$ branch, one can see that $(0 \ 1 \ 0 \ 1)$ is a feasible codeword if we can choose local codeword $\{2, 4\}$ of $c_1$ and $\{2, 4\}$ of $c_2$. However, we cannot find this feasible solution on the branch if we have only generated the local codewords $\emptyset, \{1, 2\}$ and $\{1, 4\}$ for $c_1$ and the local codeword $\emptyset$ for $c_2$. Moreover, we cannot find any other feasible solution on this branch with these limited number of local codewords. 

In such a case, the $f_2 = 1$ and $f_4 = 1$ branch is pruned by infeasibility although there is a feasible solution for LPM on the branch. 
In order to overcome this situation, we developed a column generation method based on the dual formulation. Let $P$ be the primal problem representing the RLPM and $D$ is the dual of RLPM. We first prove the following proposition:

\begin{proposition}\label{prop1.2}
$P$ is infeasible if and only if $D$ is unbounded.
\end{proposition}

\emph{Proof.} From the duality theory, we know that infeasible $P$ implies $D$ is unbounded or infeasible. We know that LPM is bounded since the variables $f_i$ and $w_{jS} \in [0, 1]$ and it is feasible since $\mathbf{0}-$codeword is a trivial solution. Then  the dual of LPM is also feasible.

$D$ being the dual of a restricted LPM, will be feasible since it contains the feasible region defined by LPM dual. This means that $D$ cannot be infeasible in any case. From here, we get $P$ is infeasible $\implies$ $D$ is unbounded.
Moreover, unbounded $D$ implies $P$ is infeasible from duality theory. As a result, we conclude that $P$ is infeasible $\iff$ $D$ is unbounded. $\square$\\

\vspace{-5mm}

At an infeasible branch, either the current $P$ is really infeasible or it occurs to be infeasible since we could not generate the columns that are necessary to construct a feasible solution. Farkas' Lemma states either $\big[ \mathbf{f} \ \  \mathbf{w} \big]  \mathbf{A=c} \text{ and } \mathbf{f \geq 0}, \mathbf{w \geq 0}$ is feasible or there is a ray $\mathbf{d}$ with $\mathbf{Ad} \leq \mathbf{0}\text{ and } \mathbf{cd}  > 0$, where $\mathbf{A}$ is the coefficient matrix for constraints and $\mathbf{c}=  \big[ \mathbf{1}  \ \ \mathbf{0} \big]$ is the right--hand--side vector of $P$. In case $P$ is infeasible, we would like to add a variable to $\mathbf{A}$ with coefficient column $\mathbf{a}^T$ with $\mathbf{ad}  > 0$ to fulfill feasibility. LP solver provides such a dual ray $\mathbf{d}$ when $P$ is infeasible. Then, we search for coefficents $\mathbf{a}$ which gives largest $\mathbf{ad}  > 0$ value. Adding columns to $P$ using dual ray obtained by solving dual Farkas system is known as Farkas pricing in the literature \cite{LD05, L10}. 

Since not all constraints (\ref{cons1}) are in $D$, $\mathbf{a}$ vector that we search is the coefficient of a dual constraint $ \mu_j - \sum_{i \in S} \tau_{ij} \leq 0$  for some $j \in C$ and $S \in \varepsilon_j$. Hence, $\mathbf{a}$ has $(m + e)-$many entries where $m$ is the number of check nodes and $e$ is the total number of edges in Tanner graph. The first $m$ entries of $\mathbf{a}$ vector are the coeffiecients for $\boldsymbol{\mu}$ variables. Then, we have zeros except a 1 for the $j$th entry. The following $e$ entries are the coefficients for $\boldsymbol{\tau}$ variables and all zero except for the --1 entries for the $j$th check node and the elements $i$ in the local codeword $S$. Let $\mathbf{d} = \big[\begin{smallmatrix} \mathbf{d^{\boldsymbol{\mu} }} \\ \mathbf{d}^{\boldsymbol{\tau}} \end{smallmatrix} \big]$, where $\mathbf{d^{\boldsymbol{\mu} }}$ and $\mathbf{d}^{\boldsymbol{\tau}}$ are the entries of $\mathbf{d}$ corresponding to the indices of the variables  $\boldsymbol{\mu}$ and $\boldsymbol{\tau}$, respectively. 
Then $\mathbf{ad} = \mathbf{d^{\boldsymbol{\mu}}} - \sum_{i \in S}\mathbf{d}^{\boldsymbol{\tau}}$ and maximizing $\mathbf{ad}$ is equivalent to maximizing $d_j^{\boldsymbol{\mu}} -  \sum_{i \in S}d_{ij}^{\boldsymbol{\tau}}$ for each check node $c_j$. 
Hence, we have to solve the following direction subproblem for each $c_j$:

%\vspace{-3mm}

$\mathbf{Direction \ Subproblem(j):}$
\vspace{-15pt}
\begin{align}
&\hspace{-3pt}\min\hspace{5pt}\sum_{i \in N(c_j)}d_{ij}^{\boldsymbol{\tau}}x_i - d_j^{\boldsymbol{\mu}}\\
&\hspace{25pt}\mbox{s.t.} \nonumber  \\  
&\hspace{25pt} \sum_{i \in N(c_j)}x_i = 2l,  \\
&\hspace{25pt} x_i \in \{0,1\}, l \in \mathbb{Z}^+.
\end{align}

We observe that the direction subproblem is actually in the same format with the column generation subproblem. Hence, on a branch we can solve the direction subproblem in $\mathcal{O}(n \log n)$ time with Algorithm 2 after replacing $\tau_{ij}$ and $\mu_j$ with $d_{ij}^{\boldsymbol{\tau}}$ and $d_j^{\boldsymbol{\mu}}$, respectively.
As a result, we can summarize our method for generating dual constraints, i.e., primal columns, with Algorithm 3.

%\vspace{-5mm}
\begin{center}
\singlespace
$
\begin{tabular}{ll}
\textbf{Algorithm 3:} (Dual constraint generation) \\
\hline
%\vspace{-4mm}\\
\textbf{Input:} An infeasible restricted primal problem, $P$\\
\hline
%\vspace{-4mm}\\
0. $isFeasible \leftarrow true$ \\
1. Solve the dual Farkas system and obtain a dual ray $\mathbf{d}$ \\
\hspace{10pt} that $D$ is unbounded.\\
2. Solve Direction Subproblem$(j)$ for each check node $j$.\\ 
\hspace{10pt} Add generated local codewords, i.e., columns, to $P$. \\
3. \textbf{If} no columns generated, \textbf{Then} conclude $P$ is infeasible.\\ 
\hspace{25pt} $isFeasible \leftarrow false$ and STOP. \\ %Prune the branch by infeasibility
4. Solve problem $P$. \\
5.  \textbf{If} $P$ is feasible, \textbf{Then} STOP. \\
6. \textbf{Else} go to Step 1.\\
7. \textbf{End If} \\
\hline
%\vspace{-4mm}\\
\textbf{Output:} $isFeasible$ \\%A feasible restricted primal problem $P$\\
%\hspace{48pt} or prune $P$ by infeasibility.\\
\hline
\end{tabular}
$
\end{center}

\subsection{Improvements to BP Algorithm} \label{Improvements} %Branch--and--Price

The general BP algorithm for IPM problem is given in Algorithm 4. BP algorithm does not implement steps $(RS)$ and $(C)$. We can improve the performance of BP algorithm in terms of solution quality and time by utilizing a new pruning rule as in Section \ref{Pruning}. BPRS method, which uses initial feasible solution generated with Random Sum (RS) heuristic given in Algorithm 5  (see Section \ref{RandomSum}), implements $(RS)$ step. In BPC method, we implement $(RS)$ step and add valid cuts (\ref{validCut}) (see Section \ref{ValidInequalities}) to RLPM in step $(C)$. 

% Additionally, we provide tight upper bounds by applying random sum heuristic in Section \ref{RandomSum}. Moreover, we make use of valid cuts given in \cite{FWK05} that eliminate fractional solutions from LPM in our BP algorithm as we explain in Section \ref{ValidInequalities}. 

\begin{center}
\singlespace
$
\begin{tabular}{ll}
\textbf{Algorithm 4:} (Solve $IPM$) \\
\hline
\vspace{-2mm}\\
\textbf{Input:} A set of feasible local codewords that constitutes $ RLPM$ \\
\hspace{35pt} ($\emptyset \in \varepsilon_j, \ \forall j$).\\ 
\hline
\vspace{-2mm}\\
0. Set $LIST = \{RLPM\}$, let $\bar{z} = \infty$ and $\underline{z} = -\infty$.\\
\emph{(RS). Apply Algorithm 5 to generate an initial feasible solution.} \\
1. \textbf{While} $LIST \neq \emptyset$ \textbf{Do}\\
2. \hspace{20pt} Select the last problem in $LIST$, say problem $P$.\\
%\hspace{34pt} /* depth--first search*/ \\
3. \hspace{20pt} Solve $P$ and obtain optimal primal $(\mathbf{f^*, w^*})$ \\ 
\hspace{34pt} and dual $(\boldsymbol{\mu}^*, \boldsymbol{\tau}^*)$ solutions with value  $\underline{z}^i$.\\
\hspace{20pt} \textit{Pruning  /* delete $P$ from the $LIST$*/}\\
4.  \hspace{20pt}  \textbf{If} $P$ is infeasible, \textbf{Then} prune by infeasibility \textbf{if} Algorithm 3 returns $false$. \\
5.   \hspace{20pt}  Go to Step 1. \\
%6.   \hspace{40pt}  \textbf{Else} Prune by infeasibility and go to Step 1. \\
6.  \hspace{20pt}  \textbf{End If} \\
7.  \hspace{20pt}  \textbf{If}  $\underline{z}^i \geq \bar{z}$, \textbf{Then} prune by bound and go to Step 1. \\
8.  \hspace{20pt}  \textbf{If} $P$ has an integer optimal solution, \textbf{Then} $\bar{z} = \underline{z}^i$,\\
\hspace{50pt} solve the subproblems with Algorithm 2. \\
9.  \hspace{40pt} \textbf{If} $\zeta_j = 0$ for all $j$,  \textbf{Then} prune by optimality, go to Step 1. \\
10.  \hspace{35pt} \textbf{Else} add the columns with $\zeta_j > 0$ to $P$, go to Step 1. \\
11.  \hspace{35pt} \textbf{End If} \\
12.  \hspace{15pt} \textbf{End If}\\
\hspace{20pt} \textit{Branching  /* add $P$ to the $LIST$*/}\\
13. \hspace{15pt} \textbf{If} $P$ has a fractional optimal solution, \\
\hspace{33pt} \textbf{Then} choose a fractional $f_i$ \\

\hspace{40pt} \textit{Left Branch}\\
14. \hspace{35pt} Let $RLPM_0$ = $P \cap \{(\mathbf{f,w}): f_i =0\}$, add  $x_i = 0$ to subproblem $j$, if $i \in N(c_j)$.\\
%\hspace{55pt} add  $x_i = 0$ to subproblem $j$, if $i \in N(c_j)$.\\
%13. \hspace{35pt} Solve the subproblems with Algorithm 2 \\
%\hspace{53pt} and add the columns with $\zeta_j > 0$ to $RLPM_0$.\\
15. \hspace{35pt} Add $RLPM_0$ to $LIST$, and go to Step 1.\\

\hspace{40pt} \textit{Right Branch}\\
16.  \hspace{35pt} Let $RLPM_1$ = $P \cap \{(\mathbf{f,w}): f_i =1\}$,  add  $x_i = 1$ to subproblem $j$, if $i \in N(c_j)$.\\
%\hspace{55pt} add  $x_i = 1$ to subproblem $j$, if $i \in N(c_j)$\\
%16. \hspace{35pt} Solve the subproblems with Algorithm 2,\\
% \hspace{53pt} and add the columns with $\zeta_j > 0$ to $RLPM_1$.\\
17. \hspace{35pt} Add $RLPM_1$ to $LIST$, and go to Step 1.\\
\emph{(C). \hspace{18pt} Apply Algorithm 6 for adding cuts (\ref{validCut}) to $RLPM$.} \\
18. \hspace{15pt} \textbf{End If}\\
19. \textbf{End While}\\
\hline
\textbf{Output:} An integral solution $(\mathbf{f^*, w^*})$ to LPM with objective value $\bar{z}$.\\
\hline
\end{tabular}
$
\end{center}

\subsubsection{A Pruning Strategy} \label{Pruning}

In a BP algorithm, we apply three pruning rules, namely prune by optimality, by infeasibility and by value dominance. We will consider an additional pruning rule that is based on the difference between the objective function values of two feasible integral solutions.

\begin{proposition}\label{prop1.6}
Let $\mathbf{f}$ be a feasible integral solution of LPM with objective function value $z$. Then, there is no feasible integral solution of LPM with objective function value in the range $(z - a, z)$ with log--likelihood objective (\ref{log_like})  where $a =  \log[(1-p)/p]$.
\end{proposition}

\emph{Proof.} From log--likelihood objective (\ref{log_like}), we can see that $z$ is an integral multiple of $a$ since $\mathbf{f}$ is integral, i.e., $z = l \cdot a$ where $l \in  \mathbb{Z}$. Let $\mathbf{f'}$ be another integral feasible solution of LPM. Then, its objective value $z'$ is also an integral multiple of $a$, say $z' = l' \cdot a$ and  where $l' \in  \mathbb{Z}$. The difference among the objectives is $z - z' = (l - l') \cdot a$.  From here, we can conclude that the nearest objective function value to $z$ can be either $z' = z + a$ or $z' = z - a$. 
Hence, there is no feasible integral solution of LPM with objective function value in the range $(z - a, z)$. $\square$

%\vspace{-5mm}

In other words, the minimum difference between two feasible integral solutions is $a$ with log--likelihood objective and 1 with Hamming distance.

\begin{proposition}\label{prop7}
Let $z$ be the optimum value of a RLPM at a branch. The branch can be pruned if $z > z_{UB} - a$, where $z_{UB}$ is the best upper bound on the IPM and $a$ is the minimum difference between two feasible integral solutions.
\end{proposition}

\emph{Proof.} A branch can be pruned by value dominance if $z > z_{UB}$. Besides, as shown in Proposition \ref{prop1.6}, there cannot be an integral feasible solution in the range $( z_{UB} - a ,  z_{UB})$. Hence, we can prune the branch if  $z > z_{UB} - a$. $\square$

\vspace{-3mm}
\subsubsection{Random Sum Heuristic} \label{RandomSum} % Feasible Solution Generation

As explained in Section \ref{ProblemDefinition}, each row of $\mathbf{G}$ is a codeword (feasible solution). We can rewrite a parity--check $\mathbf{H}$ matrix as $\mathbf{H = [A | I_{n-k}]}$ by carrying out elementary row operations under binary arithmetic. Here, $\mathbf{A}$ is a $(n-k) \times k$ binary matrix, and $\mathbf{I_{n-k}}$ is the $(n-k) \times (n-k)$ identity matrix. Then a $k \times n$ generator matrix  $\mathbf{G =  [I_{k} |  A^T]}$ can be obtained using this $\mathbf{A}$ matrix. Since one can obtain different $\mathbf{A}$ matrices, generator matrix  $\mathbf{G}$ is not unique. 

 Since $\mathbf{G}$ is a basis for the solution space of EM, any feasible solution can be written as a binary combination $\mathbf{u'}$ of the rows of $\mathbf{G}$. There are $2^k$ different  $\mathbf{u'}$  combinations, where $k$ is the number of rows of  $\mathbf{G}$. We try $ t_{max}$ random  $\mathbf{u'}$  combinations and update upper bound with the best solution found as given in Random Sum (RS) heuristic in Algorithm 5.  % We can observe that any binary combination $\mathbf{u}$ of the rows of $\mathbf{G}$ is also a feasible solution, since $\mathbf{uGH}^\textrm{T} = \mathbf{0}$ (mod 2).

% BU KISIM INTRODUCTION'A KONABİLİR.

%Another approach is to produce feasible solutions using the generator matrix $\mathbf{G}$. When a parity--check matrix $\mathbf{H}$ is given, carrying out elementary row operations under binary arithmetic, we can have a form $\mathbf{H = [A | I_{n-k}]}$ where $\mathbf{A}$ is some $(n-k) \times k$ matrix of 0's and 1's, and $\mathbf{I_{n-k}}$ is $(n-k) \times (n-k)$ identity matrix. Then a $k \times n$ generator matrix  $\mathbf{G =  [I_{k} |  A^T]}$ can be obtained using this $\mathbf{A}$ matrix. Since one can obtain different $\mathbf{A}$ matrices, the generator matrix  $\mathbf{G}$ is not unique. 

%Each of the $k$ rows of $\mathbf{G}$ is a feasible solution, since $\mathbf{GH}^\textrm{T} = \mathbf{0}$ (mod 2). From here we can see that any binary combination, $\mathbf{u}$, of the rows of $\mathbf{G}$ is also a feasible solution, since $\mathbf{uGH}^\textrm{T} = \mathbf{0}$ (mod 2). Moreover, $\mathbf{G}$ is a basis for the solution space of $\mathbf{v}\mathbf{H}^\textrm{T}=\mathbf{0}$ (mod 2). That is any feasible solution can be written as a binary combination of the rows of $\mathbf{G}$.

%--------------------------------
\vspace{-4mm}
\begin{center}
\singlespacing
$
\begin{tabular}{ll}
\textbf{Algorithm 5:} (Random Sum) \\
\hline
%\vspace{-4mm}\\
\textbf{Input:} A generator matrix $\mathbf{G}$\\ %, a received codeword  $\mathbf{\hat{y}}$
\hline
%\vspace{-4mm}\\
0. Initialize $\mathbf{z^*} = \infty, \mathbf{y^*}, t_{max}$. \\ 
1. \textbf{While} $t < t_{max}$ \\ 
2. \hspace{20pt} Randomly set $u'_i$  from \{0, 1\} for $i = 1, ..., k$.\\ 
3. \hspace{20pt} Obtain a feasible solution by $\mathbf{v} =  \mathbf{u'G}$. \\
4. \hspace{20pt} Calculate the objective function value $\mathbf{z_v}$ of solution $\mathbf{v}$. \\
5. \hspace{20pt}  \textbf{If}  $\mathbf{z_v} < \mathbf{z^*}$, \textbf{Then}  \\
6. \hspace{30pt}  $\mathbf{z^*} = \mathbf{z_v}$, $\mathbf{y^*} = \mathbf{v}$  \\
7. \hspace{20pt} \textbf{End If} \\
8. \hspace{20pt} t = t+1 \\
9. \textbf{End While} \\
\hline
%\vspace{-4mm}\\
\textbf{Output:} A feasible codeword $\mathbf{y^*}$ with objective value  $\mathbf{z^*}$. \\
\hline
\end{tabular}
$
\end{center}

%Random sum (RS) heuristic given in Algorithm 5 randomly combines the rows of generator matrix $\mathbf{G}$ and produce a new feasible solution. We pick the best known feasible solution among $k_{max}$ random trials.  
BPRS method  implements this heuristic in $(RS)$ step of Algorithm 4 in order to have tighter upper bound. Moreover, we add the columns corresponding to the best known solution to RLPM.  %In Algorithm 5, $k_{max}$ represents the maximum number of trials. % In order to speed up the row sums and objective function calculation, $BitArray$ data structure is utilized. % and in our application it is set to $k_{max} = 10000$.

\subsubsection{Valid Cuts for Fractional Solutions} \label{ValidInequalities}

As given in \cite{FWK05}, for a check node $c_j$ and all $S \subseteq N(c_j)$ with $|S|$ odd, the following inequalities are valid for IPM:

\vspace{-5mm}

\begin{equation}
\sum_{i \in N(c_j) \backslash S} f_i + \sum_{i \in S} (1 - f_i) \geq 1. \label{validCut}
\end{equation}

When we find a fractional solution in BP algorithm for RLPM, we can trim this solution if it violates inequality (\ref{validCut}) for some check node $c_j$ and odd cardinality $S \subseteq N(c_j)$. In Algorithm 6, we search for valid cuts for a given fractional solution. Adding these cuts to LPM does not affect the structure of the subproblems that are used for column generation since these cuts do not include decision variables $w_{jS}$.  Algorithm 6 generates all valid cuts (\ref{validCut}) that separate a given fractional solution. The separation algorithm runs in $\mathcal{O}(n \log n)$ time due to sorting step.

\begin{center}
\singlespacing
$
\begin{tabular}{ll}
\textbf{Algorithm 6:} (Valid Cuts) \\
\hline
\textbf{Input:} A fractional solution $\mathbf{f}$ of $RLPM$\\ 
\hline 
1. Sort $\mathbf{f}$ values in nonincreasing order. Let $\mathcal{I}_s$ be sorted indices. \\ 
2. \textbf{For Each} check node $c_j$ and odd cardinality $|S|$  \\
3. \hspace{20pt} Construct $S \subseteq N(c_j)$ using first $|S|$ neighbors of $c_j$ in $\mathcal{I}_s$. \\%and $|S|$ odd  \\
% \hspace{35pt}   \\
4. \hspace{20pt} \textbf{If} inequality (\ref{validCut}) is violated by $\mathbf{f}$ with $S$,\\
5. \hspace{20pt} \textbf{Then} add cut (\ref{validCut}) to $RLPM$.\\ 
6. \hspace{20pt} \textbf{End If} \\
7. \textbf{End For Each} \\
\hline
\textbf{Output:} Cuts added to $RLPM$. \\
\hline
\end{tabular}
$
\end{center}

We implement these cuts in BP algorithm in $(C)$ step of Algorithm 4. BP algorithm with $(RS)$ and $(C)$ steps gives rise to our BPC method.

%subproblem yapisini degistirmiyor. herhangi bir branch'te uretilen cut LPM icin de gecerli. %odd set bulma yontemi anlatılmalı ins. 

%en son BPC icin full algo verilip, step i ve ii olmazsa BP, step i olup ii olmazsa BP--IH, step ive ii oldugunda BPC methodu oluyor diye anlatıilabilir. 

\section{Computational Results} \label{ComputationalResults}

The computations have been carried out on a computer with 2.6 GHz Intel Core  i5-3230M processor and 4 GB of RAM working under Windows 10 Professional. 
We compare the performances of methods that we summarize in Table \ref{tab:som}. We implemented all methods in C\# programming language. In BP method, we apply branch--and--price algorithm in Section \ref{BranchAndPriceAlgorithm} to solve IPM formulation. BPRS method is the extension of BP method with RS heuristic. That is, we implement $(RS)$ step in Algorithm 4 in BPRS method.  BPC method is our branch--price--and--cut algorithm that uses $(RS)$ step and also add valid cuts to RLPM with $(C)$ step in Algorithm 4. We also have Exact Model Decoder (EMD) that solves EM formulation with CPLEX 12.7.1.

\begin{onehalfspace}
\begin{center}
%\footnotesize
\captionof{table}{Summary of methods}
    \label{tab:som}
\begin{tabular}{cccc} 
\hline
Method & Model & $(RS)$  & $(C)$ \\ 
\hline
BP  & IPM & --- &  --- \\
BPRS  & IPM &  $\surd$ &  ---\\
BPC  & IPM & $\surd$ & $\surd$ \\
EMD  & EM &  --- &  --- \\
\hline
\end{tabular}
\end{center}
\end{onehalfspace}

A summary of the parameters used in the computational experiments are given in Table \ref{tab:locp}. We try eight different code lengths from $n = 300$ to $n = 8400$ for three error probability $p$ levels. We randomly construct (5, 10)--regular $\mathbf{H}$ permutation code for each $n$ (see Section \ref{ProblemDefinition}). %We obtain original codeword by randomly summing the rows of $\mathbf{G}$ matrix. We generate received vector after perturbing codeword with probability $p$. For each code length $n$,  we experiment 10 random received vectors and report the average values. 
We test the quality of upper bounds obtained by RS heuristic with two different $t_{max}$ values.  In order to speed up the row sums and objective function calculation, we utilize $BitArray$ data structure for RS in Algorithm 5. We set a  time limit of 600 seconds for all methods in Table \ref{tab:som}.

\begin{onehalfspace}
\begin{center}
\captionof{table}{List of computational parameters}
    \label{tab:locp}
\begin{tabular}{l l} %p{6cm}}
\hline
\multicolumn{2}{c}{\textit{Parameters}} \\
%\textit{Parameters} & \\
\cline{1-2}
$n$  & 300, 600, 1200, 1800, 2400, 3000, 6000, 8400   \\
$p$  & 0.05, 0.07, 0.10 \\ % 0.05 (low), 0.07 (medium), 0.10 (high)
$t_{max}$ & 1000, 10000 \\
Time Limit & 600 secs \\
\hline
\end{tabular}
\end{center}
\end{onehalfspace}

Bit Error Rate (BER) is a metric in telecommunications literature to evaluate the performance of decoding algorithms. In Figure \ref{fig:diagram}, $\mathbf{v}$ is original codeword, which is received as vector $\mathbf{r}$ by digital sink. Decoder runs decoding algorithms on received vector $\mathbf{r}$ to obtain $n-$bit long decoded vector $\mathbf{y}$. BER, as given in equation (\ref{BER}), is the rate of the bits decoded different from the original codeword  $\mathbf{v}$ \cite{M03}. Note that BER = 0, if $\mathbf{v} = \mathbf{y}$. %  is the percentage of bits that are different from the original codeword

\begin{equation} \label{BER}
BER = \frac{\sum_{i=1}^{n}\mid  v_{i} -  y_{i}\mid}{n} % \times 100
\end{equation}

In our first experiment, we try  $t_{max} = 1000$ and 10000 to observe the quality of the upper bound obtained by RS heuristic (see Section \ref{RandomSum}). In Table \ref{tab:RS}, we report the average results for 30 trials for each $n$. We generate an original codeword by randomly combining the rows of $\mathbf{G}$. According to results, although it is not necessarily always the case, we can generate original codeword (BER = 0) when $t_{max} = 10000$ for all trials. Hence, we prefer to take $t_{max} = 10000$ in our BPRS and BPC methods. 

Even though we observe that RS can provide the original codeword in our experiments, we cannot calculate BER in practical applications since original codeword is unknown. That is, we cannot evaluate the solution quality of the decoded codeword of RS. On the other hand, our exact decoding methods, i.e., BP, BPRS and BPC, calculate a gap between the received vector and the decoded codeword, which is independent from the original codeword. %Hence, we cannot evaluate the solution quality of the decoded vector found by RS heuristic. In practical applications, the receiver does not know the original codeword $\mathbf{v}$. Even though RS can provide original codeword in our experiment, we cannot evaluate the solution quality of the decoded codeword in reality since BER is unkonwn.   

\newpage
\begin{onehalfspace}
\begin{center}
\footnotesize
\captionof{table}{Performance of RS heuristic}
    \label{tab:RS}
\begin{tabular}{cccccccc}
    \hline
$t_{max}$ &  \multicolumn{3}{c}{1000}  &  & \multicolumn{3}{c}{10000}  \\ \cline{2-4} \cline{6-8}
 $n$ & $z$ & BER ($\times 10^{-2}$) & CPU (secs) & & $z$ & BER ($\times 10^{-2}$) & CPU (secs) \\
    \hline
 300 & 118.5 & 41.4 & 0.03 & & 21.3 & 0 & 0.26 \\
 600 & 251.7 & 43.4 & 0.07 & & 42.4 & 0 & 0.63 \\
 1200 & 524.2 & 44.5 & 0.19 & & 84.6 & 0 & 1.83 \\
 1800 & 805.3 & 45.6 & 0.37 & & 128.3 & 0 & 3.59 \\
 2400 & 1093.4 & 46.4 & 0.62 & & 171.6 & 0 & 5.97 \\
 3000 & 1374.3 & 46.4 & 0.93 & & 215.4 & 0 & 8.45 \\
 6000 & 2816.9 & 47.4 & 3.13 & & 440.6 & 0 & 28.08 \\
 8400 & 4001.8 & 48.1 & 5.40 & & 619.7 & 0 & 53.45 \\
\hline
\end{tabular}
\end{center}
\end{onehalfspace}

%In practice, original codeword  $\mathbf{v}$ is unknown. In order to calculate BER for a decoding algorithm, one can use a feasible solution as original codeword ($\mathbf{0}-$vector is a trivial codeword) and decode it after adding random noise with probability $p$.

%Since we do not know original codeword $\mathbf{v}$, we aim in EM and IPM formulations to find the nearest codeword to the received vector $\mathbf{r}$. Hence, objective function value $z$ is the Hamming distance of the decoded vector $\mathbf{y}$ to the received vector $\mathbf{r}$ in these models.

In Table \ref{tab:EMDvsBP}, we compare the performances of EMD and BP decoders. For each code length $n$,  we experiment 10 random received vectors (in total 240 instances for three $p$ levels) and report the average values.  We give the best lower bound found by the method in column ``$z_l$," objective value of the best known solution is in column ``$z$." Although the objective functions of EM and IPM are equivalent (see Proposition \ref{prop5}), their values are different. Hence, we report the Hamming distance objective value for the solutions found by BP, BPRS, and BPC. We report the percentage difference among $z_l$ and $z$ in ``Gap." The number of instances that are solved to optimality (i.e., $z_l = z$) given in column ``\#Opt" and number of nodes evaluated in branch--and--bound tree is reported in ``\#Nodes."  BP decoder can use trivial solution of $f_i = 0$ for all $i \in V$ as an initial upper bound (see Section \ref{MathematicalFormulations}). %The corresponding objective value is in column ``$z_u^i$". 
We observe that BP method uses fewer nodes than EMD. As the code length $n$ increases, the number of nodes that both methods can evaluate decreases due to time limitation. 

When $p = 0.05$, BP is worse than EMD in terms of gap, BER and CPU. For all $p$ values, BP cannot complete the evaluation of root node for $n = 6000$ and 8400 within time limit to provide a lower bound (i.e., $z_l = 0$). As $p$ becomes 0.07 and 0.10, BP decoder provides better gap and BER figures within time limit compared to EMD. EMD solves 88 instances to optimality whereas BP finds optimum solution 61 times among 240 instances.

\newpage
\begin{onehalfspace}
\begin{center}
\tiny
\captionof{table}{Performances of EMD and BP}
    \label{tab:EMDvsBP}
\begin{tabular}{ccccccccccccccccc}
    \hline
 & &  \multicolumn{7}{c}{EMD}  &  & \multicolumn{7}{c}{BP}  \\ \cline{3-9} \cline{11-17}
  &  &  & & Gap & BER & CPU &  \multicolumn{2}{c}{\#}  & &  & & Gap & BER & CPU &  \multicolumn{2}{c}{\#}\\ \cline{8-9}   \cline{16-17} 
%$p$  & $n$ & $z_l$ & $z$&  (\%) &  ($\times 10^{-2}$) & (secs) &  Opt  & Nodes & &   $z_l$ & $z$ & (\%) & ($\times 10^{-2}$) & (secs) &  Opt  & Nodes \\
$p$  & $n$ & $z_l$ & $z$& (\%) & \hspace{-3mm} ($\times 10^{-2}$) & (secs) & \hspace{-3mm} Opt  & \hspace{-3mm} Nodes & &   $z_l$ & $z$ & (\%) & \hspace{-3mm} ($\times 10^{-2}$) & (secs) & \hspace{-3mm} Opt  & \hspace{-3mm} Nodes \\
    \hline
0.05 & 300 & 14.0 & 14.0 & 0 & 0 & 0.95 & 10 & 325.7 & &  14.0 & 14.0 & 0 & 0 & 5.33 & 10 & 0  \\
 & 600 & 28.5 & 54.9 & 8.9 & 4.9 & 66.13 & 9 & 16036.9 & & 28.1 & 28.7 & 1.6 & 0 & 82.79 & 8 & 20.7 \\
 & 1200 & 57.3 & 57.3 & 0 & 0 & 8.03 & 10 & 1232.5 & & 57.1 & 57.3 & 0.3 & 0 & 246.39 & 9 & 1 \\
 & 1800 & 87.0 & 87.0 & 0 & 0 & 0.93 & 10 & 0 & & 77.9 & 166.6 & 10.0 & 4.9 & 506.15 & 9 & 0\\
 & 2400 & 117.7 & 117.7 & 0 & 0 & 1.94 & 10 & 0 & & 105.5 & 224.9 & 10.0 & 4.9 & 527.88& 9 & 0\\
 & 3000 & 147.3 & 147.3 & 0 & 0 & 3.02 & 10 & 0 & & 103.5 & 542.7 & 30.0 & 14.6 & 553.43 & 7 & 0\\
 & 6000 & 300.9 & 565.3 & 8.9 & 5.0 & 84.74 & 9 & 83.2 & & 0 & 2963.9 & 100 & 49.4 & $time$ & 0 & 0\\
 & 8400 & 425.9 & 425.9 & 0 & 0 & 92.02 & 10 & 0 & & 0 & 4199.5 & 100 & 49.9 & $time$ & 0 & 0\\
0.07 & 300 & 18.1 & 80.0 & 42.9 & 24.5 & 301.33 & 5 & 182763.2 & & 17.2 & 62.0 & 30.1 & 0 & 303.05 & 5 & 1816.1 \\
 & 600 & 36.6 & 164.7 & 43.6 & 24.5 & 383.30 & 5 & 95563.4 & & 34.8 & 233.7 & 70.7 & 9.4 & 505.86 & 1 & 254.2 \\
 & 1200 & 71.6 & 577.8 & 87.6 & 47.8 & $time$ & 0 & 52037.7 & & 70.2 & 452.4 & 63.8 & 4.8 &  $time$ & 0 & 27.6 \\
 & 1800 & 107.6 & 891.0 & 87.9 & 50.0 & $time$ & 0 & 33363.3 & & 107.2 & 277.2 & 26.1 & 5.1 & $time$ & 0 & 4.4  \\
 & 2400 & 143.9 & 1188.6 & 87.9 & 49.8 & $time$ & 0 & 25086.2 & & 169.2 & 369.2 & 23.3 & 9.9 & $time$ & 0 & 0 \\
 & 3000 & 180.1 & 1489.1 & 87.9 & 49.8 & $time$ & 0 & 15061.8 & & 140.0 & 584.9 & 31.4 & 14.8 & 579.78 & 3 & 0\\
 & 6000 & 360.9 & 2945.1 & 87.7 & 49.1 & $time$ & 0 & 307.2 & & 0 & 2990.7 & 100 & 49.9 & $time$ & 0 & 0\\
 & 8400 & 505.4 & 4179.8 & 87.9 & 49.9 & $time$ & 0 & 9.6 & & 0 & 4168.0 & 100 & 49.5 & $time$ & 0 & 0\\
0.10 & 300 & 21.5 & 143.0 & 84.9 & 49.1 & $time$ & 0 & 430245.8 & & 19.6 & 156.4 & 87.5 & 47.3 & $time$ & 0 & 5201.0 \\
 & 600 & 40.5 & 291.9 & 86.1 & 49.0 & $time$ & 0 & 161875.3 & & 38.9 & 297.3 & 86.9 & 50.0 & $time$ & 0 & 235.4 \\
 & 1200 & 78.6 & 586.6 & 86.6 & 49.7 & $time$ & 0 & 60389.7 & & 77.4 & 537.8 & 81.5 & 14.6 & $time$ & 0 & 24.6 \\
 & 1800 & 118.2 & 882.2 & 86.6 & 49.6 & $time$ & 0 & 32469.2 & & 118.2 & 744.2 & 76.0 & 14.8 & $time$ & 0 & 4.0 \\
 & 2400 & 156.9 & 1189.2 & 86.8 & 49.9 & $time$ & 0 & 24190.8 & & 190.1 & 801.8 & 61.3 & 14.6 & $time$ & 0 & 0 \\
 & 3000 & 196.1 & 1470.5 & 86.7 & 49.1 & $time$ & 0 & 12114.8 & & 144.3 & 774.5 & 50.7 & 20.2 & $time$ & 0 & 0 \\
 & 6000 & 391.7 & 2964.7 & 86.8 & 49.2 & $time$ & 0 & 244.9 & & 0 & 2972.9 & 100 & 50.0 & $time$ & 0 & 0 \\
 & 8400 & 548.0 & 4167.2 & 86.8 & 49.6 & $time$ & 0 & 0.3 & & 0 & 4179.2 & 100 & 49.6 & $time$ & 0 & 0 \\
\hline
\end{tabular}
\end{center}
\end{onehalfspace}

Table \ref{tab:BPRSvsBPC} summarizes the results for BPRS and BPC methods. In both methods, we provide an inital solution with RS heuristic. We report the number of valid cuts (\ref{validCut}) used by BPC method in column ``\#Cuts." BPRS and BPC can find original codeword (i.e., BER = 0) for all instances either with RS heuristic or improving the solution of RS with BP algorithm. The number of cases  out of 240 instances solved to optimality for BPRS and BPC are 142 and 161, respectively. As we improve BP algorithm to BPRS and BPC, we observe better gap, BER, \#Opt values and fewer nodes. Moreover, BPC gives better figures for these performance metrics compared with EMD.

Table \ref{tab:BPRSvsBPC} shows that our algorithms can solve more instances to optimality as code length $n$ increases for $p = 0.07$ and $0.10$ which is somewhat counter intuitive. For example, we can solve 10 instances to optimality when $n \geq 6000$ for these $p$ values. This is due to $\mathbf{H}$ codes and properties of LPM formulation. Randomly constructed permutation codes have fewer cycles in their Tanner graph representations as the dimension of the code gets larger \cite{LMT+17}. When Tanner graph is cycle--free, any optimum solution of LPM is integral as noted in Feldman \emph{et al.} \cite{FWK05}. Hence, as code length $n$ increases, we have Tanner graph with fewer cycles, which results in a better LP lower bound at the root node. % of BP algorithm.
 This is not apparent for BP, since we cannot complete root node evaluation due to time limit for $n \geq 6000$. We succeeded for BPRS and BPC methods with the help of RS heuristic.

%We either get an initial upper bound with BER = 0 and prove that it is optimal solution with BP or we get an initial upper bound with BER $>$ 0 and improve that solution to optimality with BP algorithm.

\begin{onehalfspace}
\begin{center}
\tiny
\captionof{table}{Performances of BPRS and BPC}
    \label{tab:BPRSvsBPC}
\begin{tabulary}{6cm}{cccccccccccccccccc}
    \hline
 & &  \multicolumn{7}{c}{BPRS}  &  & \multicolumn{8}{c}{BPC}  \\ \cline{3-9} \cline{11-18}
  &  &  & & Gap & \hspace{-3mm} BER & CPU &  \multicolumn{2}{c}{\#}  & &  & & Gap & \hspace{-3mm} BER & CPU &  \multicolumn{3}{c}{\#}\\ \cline{8-9}   \cline{16-18} 
$p$  & $n$ & $z_l$ & $z$& (\%) & \hspace{-3mm} ($\times 10^{-2}$) & (secs) & \hspace{-3mm} Opt  & \hspace{-3mm} Nodes & &   $z_l$ & $z$ & (\%) & \hspace{-3mm} ($\times 10^{-2}$) & (secs) & \hspace{-3mm} Opt  & \hspace{-3mm} Nodes & \hspace{-3mm} Cuts\\
    \hline
0.05 & 300 & 14.0 & 14.0 & 0 & 0 & 5.61 & 10 & 0 & & 14.0 & 14.0 & 0 & 0 & 124.05 & 10 &  0 &  47.1 \\
 & 600 & 28.1 & 28.7 & 1.6 & 0 & 87.93 & 8 & 18.8 & & 28.6 & 28.7 & 0.4 & 0 & 146.01 & 9 & 0 & 83.2 \\
 & 1200 & 57.1 & 57.3 & 0.3 & 0 & 272.99 & 9 & 1.2 & & 57.3 & 57.3 & 0 & 0 & 342.86 & 10 & 0 & 97.0 \\
 & 1800 & 87.0 & 87.0 & 0 & 0 & 405.58 & 10 & 0 & & 87.0 & 87.0 &  0 & 0 & 413.26 & 10 & 0 & 0 \\
 & 2400 & 117.7 & 117.7 & 0 & 0 & 520.88 & 10 & 0 & & 117.7 & 117.7 & 0 & 0 & 534.42 & 10 & 0 & 0 \\
 & 3000 & 147.3 & 147.3 &  0 & 0 & 560.27 & 10 & 0 & & 147.3 & 147.3 & 0 & 0 & 551.42 & 10 & 0 & 0 \\
 & 6000 & 301.7 & 301.7 & 0 & 0 & 550.75 & 10 & 0 & & 301.7 & 301.7 & 0 & 0 & 535.01 & 10 & 0 & 0 \\
 & 8400 & 425.9 & 425.9 & 0 & 0 & 529.42 & 10 & 0 & & 425.9 & 425.9 & 0 & 0 & 521.18 & 10 & 0 & 0 \\
0.07 & 300 & 17.6 & 19.4 &  7.7 & 0 & 242.21 & 6 & 307.8 & & 18.8 & 19.4 & 2.5 & 0 & 421.27 & 6 & 0 & 153.4 \\
 & 600 & 35.2 & 39.3 & 9.6 & 0 & 360.41 & 3 & 119.5 & & 38.5 & 39.3 & 1.9 & 0 & 543.37 & 5 & 0 & 299.8  \\
 & 1200 & 70.3 & 79.0 &10.8 & 0 & $time$ & 0 & 27.7 & & 77.3 & 79.0 & 2.1 & 0 & 598.01 & 3 & 0 & 752.4 \\
 & 1800 & 107.2 & 120.8 & 11.0 & 0 & 587.65 & 1 & 3.9 & & 119.1 & 120.8 & 1.4 & 0 & 580.13 & 4 & 0 & 1113.3 \\ 
 & 2400 & 151.7 & 162.2 &6.3 & 0 & 539.93 & 3 & 0 & & 159.7 & 162.2 & 1.5 & 0 & 550.47 & 4 & 0 & 806.1 \\
 & 3000 & 203.0 & 204.1 & 0.5 & 0 & 576.46 & 8 & 0 & & 203.6 & 204.1 & 0.2 & 0 & 546.35 & 9 & 0 & 120.2 \\
 & 6000 & 420.1 & 420.1 & 0 & 0 & 535.16 & 10 & 0 & & 420.1 & 420.1 & 0 & 0 & 551.19 & 10 & 0 & 0 \\
 & 8400 & 591.2 & 591.2 & 0 & 0 & 544.68 & 10 & 0 & & 591.2 & 591.2 & 0 & 0 & 563.28 & 10 & 0 & 0 \\
0.10 & 300 & 19.6 & 30.6 & 34.4 & 0 & $time$ & 0 & 754.6 & & 28.7 & 30.6 & 5.7 & 0 & 593.01 & 4 &  0 & 68.8  \\
 & 600 & 38.9 & 59.3 & 33.9 & 0 &  $time$ & 0 & 206.7 & & 57.0 & 59.3 & 3.9 & 0 & 588.14 & 3 &  0 & 140.4 \\
 & 1200 & 77.4 & 117.6 & 34.1 & 0 & $time$ & 0 & 25.8 & & 110.3 & 117.6 & 6.1 & 0 & $time$ & 0 &  0 & 359.3  \\
 & 1800 & 117.4 & 177.2 &  33.6 & 0 & $time$ & 0 & 3.4 & & 171.1 & 177.2 & 3.4 & 0 & 592.98 & 1 &  0 & 422.3  \\
 & 2400 & 171.6 & 234.8 & 26.9 & 0 & $time$ & 0 & 0 & & 225.0 & 234.8 &  4.2 & 0 & $time$ & 0 &  0 & 608.0   \\
 & 3000 & 268.4 & 294.9 & 8.9 & 0 & 565.83 & 4 & 0 & & 285.2 & 294.9 & 3.3 & 0 & 521.89 & 3 &  0 & 459.0 \\
 & 6000 & 599.9 & 599.9 & 0 & 0 & 543.84 & 10 & 0 & & 599.9 & 599.9 & 0 & 0 & 545.63 & 10 &  0 & 0  \\
 & 8400 & 842.0 & 842.0 & 0 & 0 & 558.98 & 10 & 0 & & 842.0 & 842.0 & 0 & 0 & 531.61 & 10 &  0 & 0   \\
\hline
\end{tabulary}
\end{center}
\end{onehalfspace}

In practical applications, decoding of a received vector is done with iterative algorithms, such as Gallager A given in Algorithm 7, with low complexity \cite{L05}. In Gallager A, $v_i$ is incident to $d_i$ many check nodes on Tanner graph and $u_i$ many of them are unsatisfied. A bit $i$ is candidate to be flipped, if  $u_i > d_i / 2$. At each iteration, Gallager A flips only a candidate bit $i$ with largest $u_i$ value. %Gallager A guarantees to decrease the number of unsatisfied check nodes, since it flips only one bit at each iteration.

\begin{onehalfspace}
\begin{center}
\footnotesize
$
\begin{tabular}{ll}
\textbf{Algorithm 7:} (Gallager A) \\
\hline
\vspace{-4mm}\\
\textbf{Input:} Received vector, $\mathbf{r}$\\
\hline
\vspace{-4mm}\\
1.  Calculate all parity--check equations \\
2. \textbf{If} all check nodes are satisfied, \textbf{Then} STOP.\\ 
3. \textbf{Else} Calculate the number of all unsatisfied parity--check \\
\hspace{40pt} equations for each received bit, $u_i$ for bit $i$. \\
4. \hspace{20pt} Let $l = \text{argmax}_i\{u_i\}$. \textbf{If} $u_l > d_l / 2$, \textbf{Then} flip bit $l$.\\
5. \textbf{End If} \\
6. \textbf{If} stopping condition is satisfied, \textbf{Then} STOP.\\ 
7. \textbf{Else} Go to Step 1.\\
8. \textbf{End If} \\
\hline
\vspace{-4mm}\\
\textbf{Output:} A feasible decoded codeword, or no solution\\
\hline
\end{tabular}
$
\end{center}
\end{onehalfspace}
\vspace{5mm}

In our final experiment, we compare our proposed decoding algorithms with Gallager A algorithm. Iterative algorithms may get stuck and terminate with no conclusion when there is a cycle in Tanner graph \cite{SPT14}. To avoid such a situation, we take the stopping criterion in Algorithm 7 as 500 iterations. Note that this may result in ending with an infeasible solution when Gallager A terminates. 

\vspace{-4mm}
\begin{onehalfspace}
\begin{center}
\footnotesize
\captionof{table}{Performance of Gallager A}
    \label{tab:GallagerA}
\begin{tabular}{cccccc}
    \hline
$p$ & $n$ & $z$ & BER ($\times 10^{-2}$) & CPU (secs) & \#Feas \\
    \hline
0.05 & 300 & 14.0 & 4.5 & 0.63 & 4 \\
 & 600 & 26.7 & 4.2 & 2.52 & 0 \\
 & 1200 & 52.6 & 3.8 & 12.50 & 2 \\
 & 1800 & 84.6 & 3.4 & 31.77 & 3 \\
 & 2400 & 116.1 & 4.1 & 54.26 & 0 \\
 & 3000 & 142.5 & 3.8 & 84.53 & 2 \\
 & 6000 & 283.7 & 4.5 & 367.67 & 0 \\
 & 8400 & 405.3 & 4.9 & 882.95 & 0 \\
0.07 & 300 & 18.3 & 8.6 & 0.63 & 1 \\
 & 600 & 36.1 & 8.7 & 2.65 & 0 \\
 & 1200 & 71.0 & 9.2 & 12.52 & 0 \\
 & 1800 & 113.7 & 9.5 & 30.08 & 0 \\
 & 2400 & 147.2 & 9.6 & 54.50 & 0 \\
 & 3000 & 189.1 & 9.7 & 84.18 & 0 \\
 & 6000 & 377.5 & 10.2 & 363.77 & 0 \\
 & 8400 & 486.0 & 9.7 & 697.95 & 0 \\
0.10 & 300 & 20.1 & 14.1 & 0.62 & 0 \\
 & 600 & 43.6 & 14.6 & 2.56 & 0 \\
 & 1200 & 84.9 & 14.4 & 12.15 & 0 \\
 & 1800 & 127.0 & 13.9 & 29.86 & 0 \\
 & 2400 & 172.0 & 14.3 & 54.22 & 0 \\
 & 3000 & 216.7 & 14.3 & 84.50 & 0 \\
 & 6000 & 434.4 & 14.7 & 364.64 & 0 \\
 & 8400 & 492.2 & 13.6 & 723.52 & 0  \\
\hline
\end{tabular}
\end{center}
\end{onehalfspace}

In Table \ref{tab:GallagerA}, we summarize the results of Gallager A algorithm.  Gallager A finds a feasible solution for ``\#Feas"  instances among 10 instances. One can observe that $z$ values in Table \ref{tab:GallagerA} are smaller than $z_l$ values of BPC in Table \ref{tab:BPRSvsBPC} due to infeasible solutions found by Gallager A. BPC method can find original codeword (i.e., BER = 0) for all instances, whereas Gallager A finds infeasible vectors that are far away from the original codeword (i.e., BER $>$ 0) in most cases. Gallager A can decode to original codeword only for 12 among 240 instances.

Computation times reported in the ``CPU (secs)" columns of Tables  \ref{tab:BPRSvsBPC} and \ref{tab:GallagerA} indicate that Gallager A is faster than our BPC method. On the other hand, BPC method can find higher quality solutions in the expense of decoding duration. The applications such as TV broadcasting and video streaming, in which the decoding latency is the key issue, implement fast decoding algorithms as Gallager A. However, there are cases such as deep space communications, that we cannot reobtain the information from the digital source. For such applications, high quality decoding is important instead of decoding speed. Our BPC method is a candidate decoder thanks to its high deoding quality for such communication systems.

\section{Conclusions} \label{Conclusions}

In this study, we focus on decoding algorithms that correct the errors in received vector using LDPC codes for digital communication systems. We consider a mathematical formulation from the literature and propose a branch--and--price (BP) algorithm for its solution. We improve the error correction capability of our BP algorithm by providing tight upper bounds with random sum (RS) heuristic and introducing valid cuts to mathematical formulation. These enhancements give rise to our branch--and--price--random--sum (BPRS) and branch--price--and--cut (BPC) methods. 

Our computational experiments show that our BPC method outperforms exact model decoder (EMD), which makes use of commercial solver CPLEX 12.7.1, in terms of gap, BER and number of instances solved to optimality.  Moreover, BPC method can find near optimum feasible solutions, whereas practically used iterative decoder Gallager A algorithm terminate with infeasible solutions far from the original codeword in most of the cases having high error rates. 

Our BPC decoder can contribute to the construction of reliable digital communication systems with its high error correction capability. In particular, BPC can be used for the critical applications, such as NASA's Mission Cassini, in which we receive the information only once. In such settings, solution quality is crucial instead of decoding latency. Considering decoding is an online problem, faster decoders are desired. Hence, improving the solution time of BPC method can be a future research.

\vspace{-5mm}

\section*{Acknowledgments} 
This research has been supported by the Turkish Scientific and Technological Research Council with grant no 113M499.

\begin{comment}
Dilekce duası
Bismillahirrahmanirrahim. Vaadallahüs sabirine nasran. Ve kaddere limen yetevekkele aleyhi ecren. Ve şereha limen yüfevvizu ileyhi sadren. Fe inne maalusri yüsren, inne maal usri yüsra. İnne kitabel ebrari lefi illiyyin. Ve ma edrake ma illiyyun. Kitabün merkum. Yeşhedühül mukarrebun. 

time invariantta m model 1 modele iner. performansı dusuk implementation kolay.

\end{comment}

\end{document}